\documentclass[final,1pt,times,twocolumn]{elsarticle}

\usepackage{amssymb}
\usepackage{amsmath}

\usepackage{subcaption}

\newcommand{\be}{\begin{eqnarray}}
\newcommand{\ee}{\end{eqnarray}}
\newcommand{\beq}{\begin{equation}}
\newcommand{\eeq}{\end{equation}}
\newcommand{\bemul}{\begin{multline}}
\newcommand{\eemul}{\end{multline}}

\journal{Physics Letters B}
\begin{document}
\begin{frontmatter}
\title{A choked jet in SN 2023ixf?}
\author[inst1]{Matías M. Reynoso}
\affiliation[inst1]{organization={Instituto de Investigaciones Físicas de Mar del Plata (IFIMAR – CONICET), and Departamento de Física, Facultad de Ciencias Exactas y Naturales, Universidad Nacional de Mar del Plata},
            addressline={Funes 3350}, 
            city={Mar del Plata},
            postcode={7600}, 
            state={Provincia de Buenos Aires},
            country={Argentina}}
\begin{abstract}
It has been proposed that core-collapse supernovae (CCSNe) can take place along with the generation of a jet that fails to emerge from the stellar envelope of the progenitor star, i.e., a choked jet. Although the fraction of CCSNe that harbour such jets is unknown, it remains as an interesting possibility that can give rise to the production of high-energy neutrinos. In this work, we focus on the particular case of the recent type II supernova, SN 2023ixf, the closest of its class in the last decade. We describe the particle distributions of protons, pions, and muons in a putative jet applying a simple model to account for the relevant interactions, which are synchrotron cooling and interactions with the soft photon field in the ambient. After evaluating the produced fluence for different values of the viewing angle $i_{\rm j}$ with respect to the jet axis, and comparing with the upper bound by IceCube, we conclude that the generation of a choked jet in SN 2023ixf can not be ruled out. Specifically, for typical jet Lorentz factors of the jet $\Gamma=100$, the jet could have been produced with an half-opening angle $\theta_{\rm op}=0.2\,{\rm rad}$ but for a viewing angle  $i_{\rm j}\gtrsim \theta_{\rm op}$, no significant Doppler boosting would take place along this direction. Therefore, the choked jet scenario in SN 2023ixf still remains compatible with observations provided our line of sight corresponds to an off-axis view of the jet.
\end{abstract}


\end{frontmatter}



\section{Introduction} \label{sec:intro}
The observation of neutrinos from astrophysical sources potentially opens a remarkable window to study different aspects of the physics governing such objects. In the particular case of core-collapse supernovae (CCSNe), the gravitational collapse of massive stars triggers the formation of a proto-neutron star in the innermost parts of the core. As the rest of the surrounding material bounces back and produces an outward expanding shock, neutrinos are generated carrying away a huge amount of energy, $\sim 10^{53}{\rm erg}$ \citep{colgate1966}, i.e., 100 times more than what is required to power a SN. This picture was confirmed by the detection of MeV neutrinos from the type II SN 1987a \citep{bionta1987,hirata1987}.
In this type of SNe, a stellar envelope of hydrogen is retained, as has been inferred through observations of the corresponding spectral lines \citep[e.g.][]{filippenko1997}. Therefore, if a jet is launched by the central compact object, it can be stalled inside this envelope \citep{macfadyen2001}. 
Similarly to the proposed models for usual GRBs, choked jets can present internal shocks (IS) and accelerate protons that may produce neutrinos via $p\gamma$ interactions, as has largely been studied under different considerations \citep{meszaros2001,razzaque2004,murase2013,senno2016,he2018,fasano2021,reynoso2023}. 

The contribution of choked jets in CCSNe to the diffuse neutrino flux observed by IceCube could be significant \citep{chang2022}, although recent studies obtain no significant correlation between observed SNe and neutrino events \citep{abbasi2023}. A key factor in this sense may be the fraction of such sources presenting jets, which is still uncertain \citep{piran2017}.
We note that these CCSNe can also generate high energy neutrinos at later stages, where the SN ejecta interacts with the circumstellar matter at larger distances from the core \citep[e.g.][]{murase2018,kheirandish2023}. While in that scenario no jet is required, here we still explore the possibility of the generation of a choked jet in CCSNe, particularly in the case of the recent SN 2023ixf. This is the closest type II SN in the last decade \citep{itagaki2023,perley2023}, with M101 as its host galaxy at a distance $d_L=6.85\, {\rm Mpc}$ from Earth \citep{croxall2016}. It has been the subject of many studies and is still the target for observations \citep[e.g.][]{yamanaka2023,bostroem2023,bersten2024,chandra2023}. In particular, IceCube has briefly reported upper bounds for a neutrino signal within a time window of $\pm 2$ days since its discovery in the optical band \citep{thwaites2023}. This has been considered in Ref. \cite{guetta2023} to assess the plausibility of high energy neutrino production neutrinos by a choked jet in this source.

In the present work, we examine in detail the possibility that a choked jet may have indeed been generated in SN 2023ixf, assuming that protons can be accelerated by IS in such a jet. We apply the model discussed in Ref. \cite{reynoso2023}, which taking into account all the relevant cooling processes, allows to obtain the proton, pion, and muon distributions as solutions to stationary transport equations. Adopting typical values for the physical parameters involved, we obtain the neutrino fluence for different values of the viewing angle $i_{\rm j}$. We find that the non-detection by IceCube can be consistent with the choked jet scenario in SN 2023ixf provided that the viewing angle is larger that the jet half-opening angle, i.e., we would be seen an off-axis view of the system.  

The rest of the manuscript is organized as follows. In Section \ref{sec:model}, we describe the adopted model to characterize the neutrino emission by ISs in choked jets of SNe. In Section \ref{sec:nuflux} we present the resulting neutrino fluxes expected for the particular case of SN 2023ixf under different assumptions for the set of key physical parameters, such as the jet power, magnetic field, injection index of the accelerated particles, and viewing angle. Finally, in Section \ref{sec:discussion} we conclude with a discussion to remark that a choked jet can have been produced in SN 2023ixf, and therefore its role in type II SNe remains to be further explored.

\section{Choked jets in type II SNe} \label{sec:model}
The occurrence of jets in CCSNe has been signalled as a possible natural outcome in the context of collapsing massive stars.
{Generically, jet production is more naturally favoured in massive stars that have lost their hydrogen layer \citep{woosley2006} and present a fast rotating core with a high magnetic field \citep{heger2005}. These conditions are particularly necessary when trying to understand the most energetic events associated to long gamma-ray bursts (GRBs), or hypernovae \citep[e.g.][]{obergaulinger2022}. In the case of type II SNe, the collapsing supergiant progenitors are in principle not typically expected to be fast rotating or highly magnetized. However, the generation of a jet in some of such systems may not be ruled out completely, as has been considered in different studies \citep[e.g.][]{couch2009,smith2012,soker2022}. Specifically, one possibility first envisaged in \citep{macfadyen2001} may be realized in an ongoing SN explosion through a fallback of matter taking place minutes to hours after the formation of a neutron star in the core of a supergiant star. A black hole is created at that point and an hyperdense accretion disk is formed around it with the conditions to allow for jet launching through magnetohydrodynamic processes \citep{blandfordznajek1977}. Such a jet would be choked, as it would be stalled in the supergiant star envelope as mentioned above.}

We therefore adopt a similar scenario for choked jets as those explored in several works since 2001 \citep[e.g.,][]{macfadyen2001,meszaros2001,murase2013,senno2016,he2018,fasano2021}. {In particular,  we focus on the case of jets propagating inside a red supergiant star as progenitor of a type II SN \cite{he2018, fasano2021}. As anticipated, we apply the treatment discussed in detail in Ref. \citep{reynoso2023}, in order to account for the cooling processes of the particle populations involved in the neutrino production process. Here we summarize the main assumptions and outline the method of calculation. The Lorentz factor of the jet is considered to be $\Gamma\sim 10-300$ and its half-opening angle $\theta_{\rm op}\approx 0.2 {\rm rad}$. The jet power $L_{\rm j}$ corresponds to an isotropic equivalent power 
\be 
L_{0}= 2L_{\rm j}/(1-\cos\theta_{\rm op})= 10^{49-50}{\rm erg \, s^{-1}}. 
\ee
Setting a variability timescale $\delta t_{-2}=\delta t/(0.01 {\rm s})$, the distance from the central source to the IS position in the jet is
\be
r_{\rm is}= 2\Gamma^2c\delta t\simeq 6\times 10^{12}{\rm cm}\ {\delta t_{-2}} \Gamma^2_{2},
\ee
where $\Gamma_2=\Gamma/100$. The comoving number density of cold protons is given by
\be
n'_{\rm j}= \frac{L_{0}}{4\pi\Gamma^2 r^2_{\rm is} m_p c^3} \simeq 4.9\times 10^{11}{\rm cm^{-3}}L_{0, 50}\Gamma_2^{-6}\delta t_{-2}^{-2}, 
\ee 
where $L_{0,50}=\frac{L_0}{10^{50}{\rm erg/s}}.$
Likewise, the magnetic energy density is assumed to be a fraction $\epsilon_{B}=0.1$ of the kinetic energy density, so that 
\be
B'=\sqrt{\frac{2 \,\epsilon_B L_{0}}{\Gamma^2 r^2_{\rm is}  c}}= 4.3\times 10^{4}{\rm G} \ \epsilon_{B,-1}^{{1}/{2}}L_{0,50}^{{1}/{2}} \Gamma_{2}^{-2} \delta t_{-2}^{-1}.
\ee \label{Bmag}
The jet propagates through the red giant star and it is assumed to stall within the hydrogen envelope, which extends up to a radius $r_{\rm ext}\approx 10^{13.5}{\rm cm}$ \citep[e.g.][]{guetta2023,kilpatrick2023}. As the jet is stalled, a forward shock and a reverse shock develop \cite{bromberg2011} and \cite{mizuta2013}. The jet head is the part of the jet that is affected by the reverse shock, with a Lorentz factor $\Gamma_{\rm h}\simeq 1$ {\citep[e.g.,][]{he2018},} at a position given by \citep[e.g.,][]{mizuta2013,murase2013}
\be 
  r_{\rm h}\simeq 1.3\times 10^{13}{\rm cm} \, t_{\rm j,3}^{3/5}L_{0,50}^{1/5}\left(\frac{\theta_{\rm op}}{0.2}\right)^{-4/5}\rho_{{\rm ext},-7}^{-1/5}.
\ee 
Here, the jet is assumed to be active during a time $t_{\rm j}$, with $t_{\rm j,3}=t_{\rm j}/(10^3\,{\rm s})$, and the density of the envelope is typically $\rho_{\rm ext,-7}=\rho_{\rm ext}/(10^{-7}{\rm g \ cm^{-3}})$. The jet becomes collimated at and keeps a constant radius up to the position $r_{\rm h}$, where it is finally stopped. Collimation shocks are expected to be generated at a distance from the central source given by
\citep{bromberg2011,mizuta2013}}
\be 
r_{\rm cs}\simeq 8.2\times 10^{12}{\rm cm} \,t_{\rm j,3}^{2/5}{L_{0,50}}^{3/10} \rho_{\rm ext,-7}^{-3/10}.
\ee

\begin{table*}[h!]
{\small 
\caption{Parameter sets of the CGRB model}\label{table:params}
\centering                                      
\begin{tabular}{l c c c c c}          
\hline                     
parameter &  description &   set $A_1$ & set $A_2$ & set $B_1$ & set $B_2$ \\    
\hline                                   
    $\delta t {\rm[s]}$ & variability timescale  & $10^{-2}$& $10^{-2}$& $10^{-3}$& $10^{-3}$ \\
    $L_0$[erg/s]  &  isotropic power  & $10^{49}$ & $10^{50}$& $10^{49}$ & $10^{50}$ \\      
    $t_{\rm j}$[s] &  jet duration&  $10^4$& $10^3$ &  $10^4$& $10^3$ \\
    $\epsilon_{\rm rel}$ & ratio $L_p/L_0$ & 0.1 & 0.1 & 0.1 & 0.1 \\
$\epsilon_B$  & magnetic-to-kinetic energy ratio & $0.1$& $0.1$& $0.1$& $0.1$  \\    
    $\Gamma$ & Lorentz factor of IS region & $(30..130)$ &  $(50..115)$ &  $(50..350)$ &  $(75..500)$ \\
   
\hline
\end{tabular}
}
\end{table*}

\begin{figure*}[]                   
\centering
\  \centering
    \begin{subfigure}[t]{0.49\textwidth}
        \centering                          
        \includegraphics[width=0.5\linewidth,trim= 180 30 190 35]{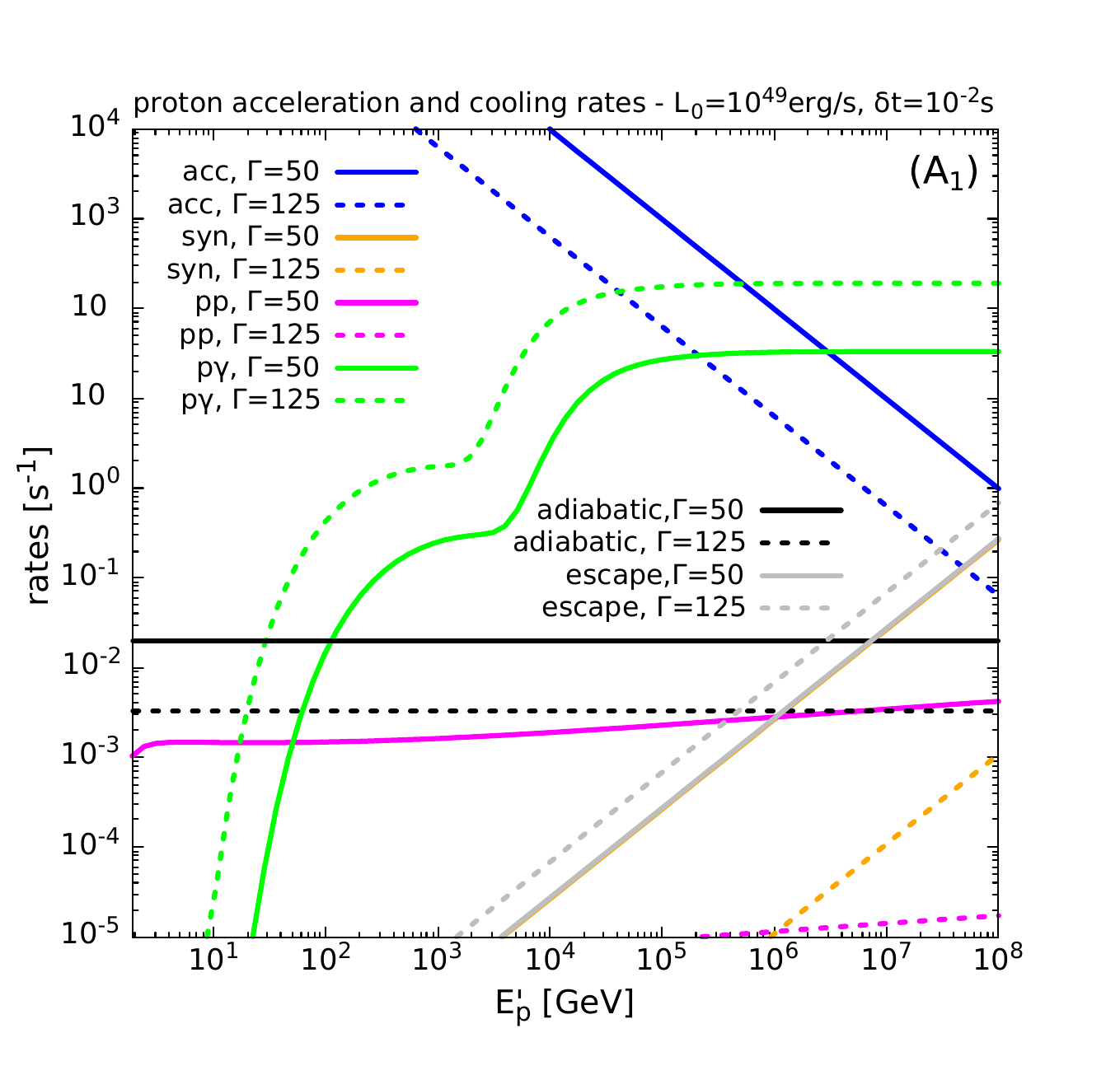} 
    \end{subfigure}
    \hfill
    \begin{subfigure}[t]{0.49\textwidth}
        \centering
        \includegraphics[width=0.5\linewidth,trim= 180 30 190 35]{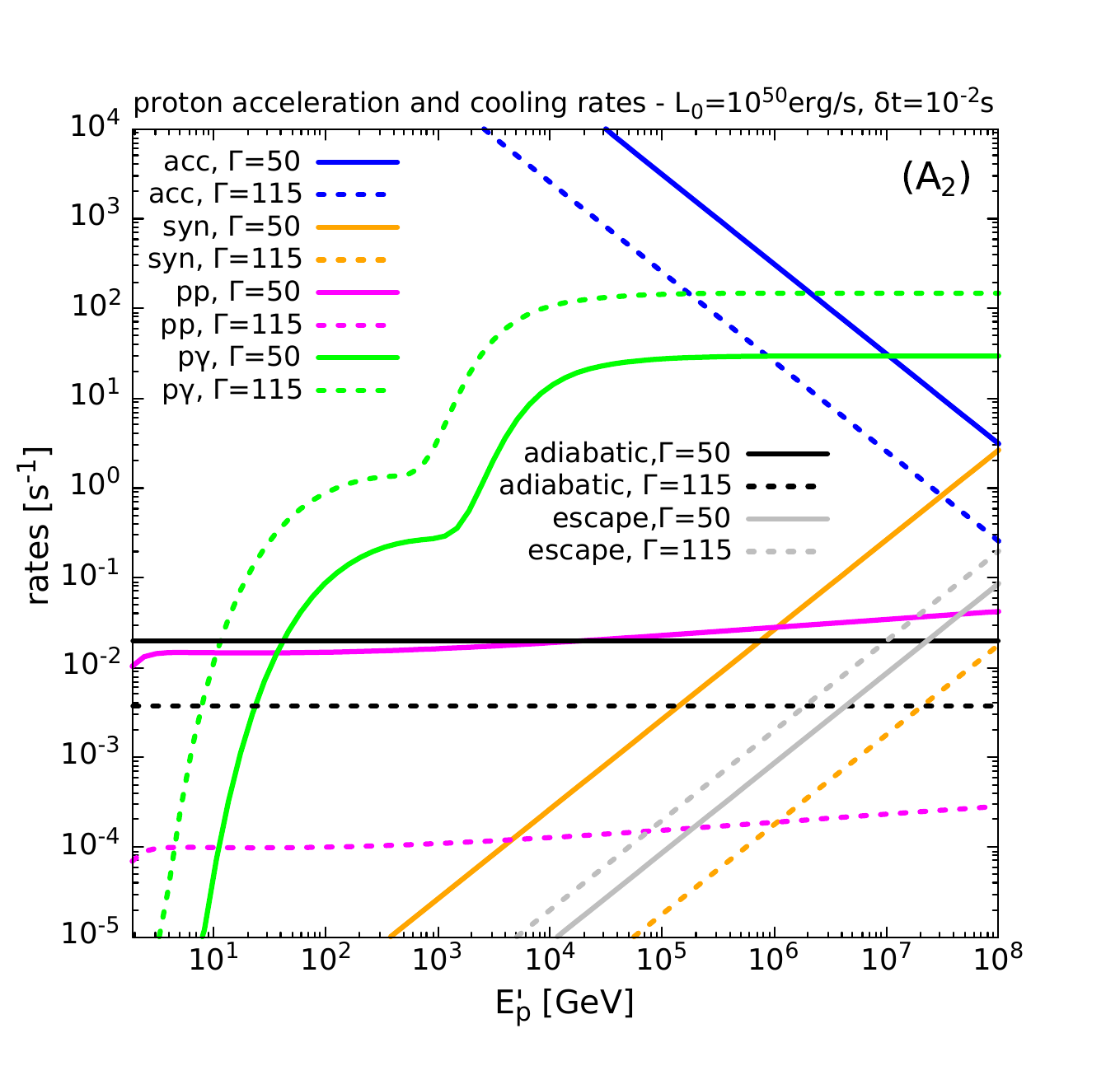} 
    \end{subfigure}
    \begin{subfigure}[t]{0.49\textwidth}
        \centering                     
        \includegraphics[width=0.5\linewidth,trim= 180 30 190 35]{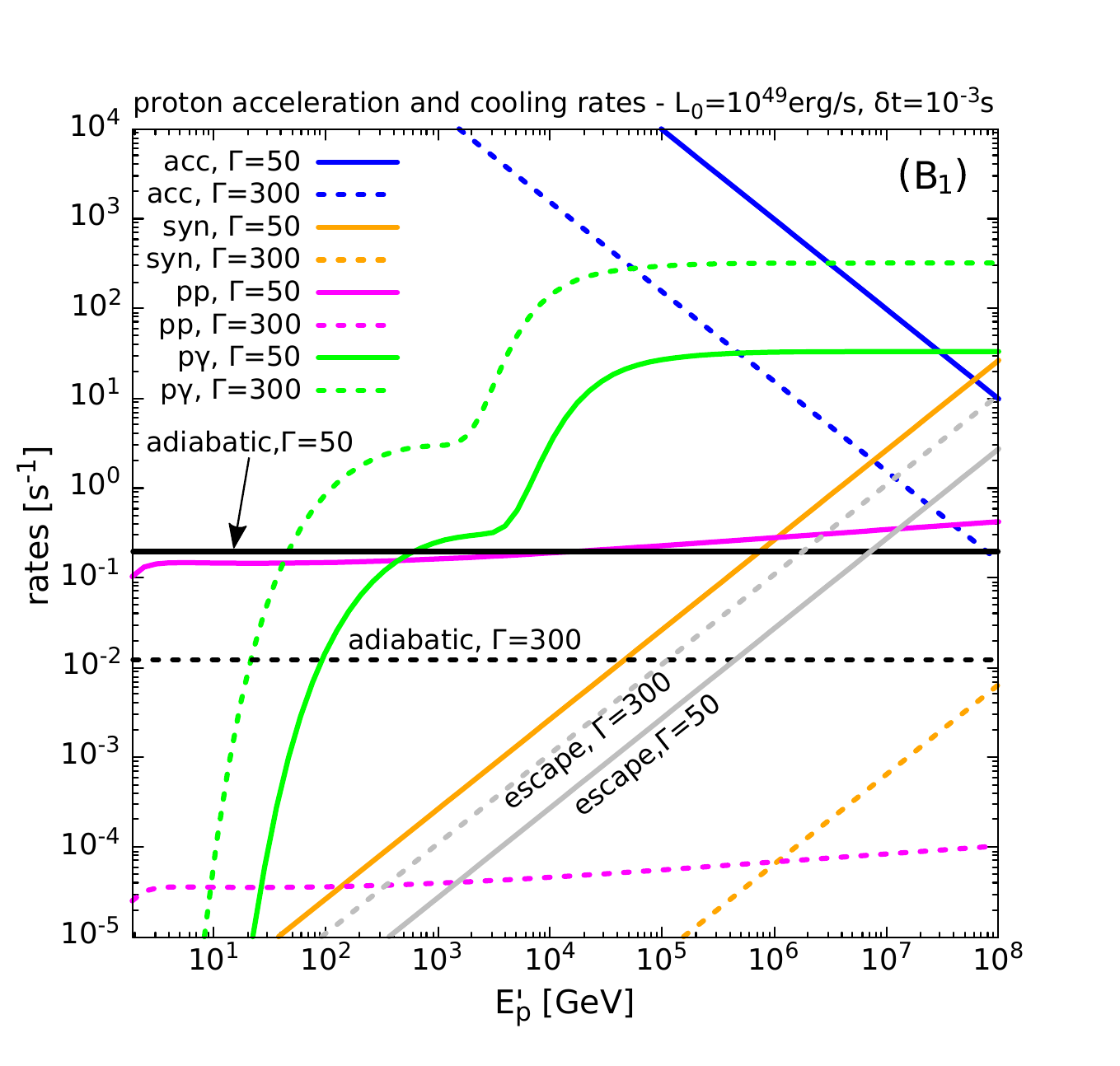} 
    \end{subfigure}
    \hfill
    \begin{subfigure}[t]{0.49\textwidth}
        \centering
        \includegraphics[width=0.5\linewidth,trim= 180 30 190 30]{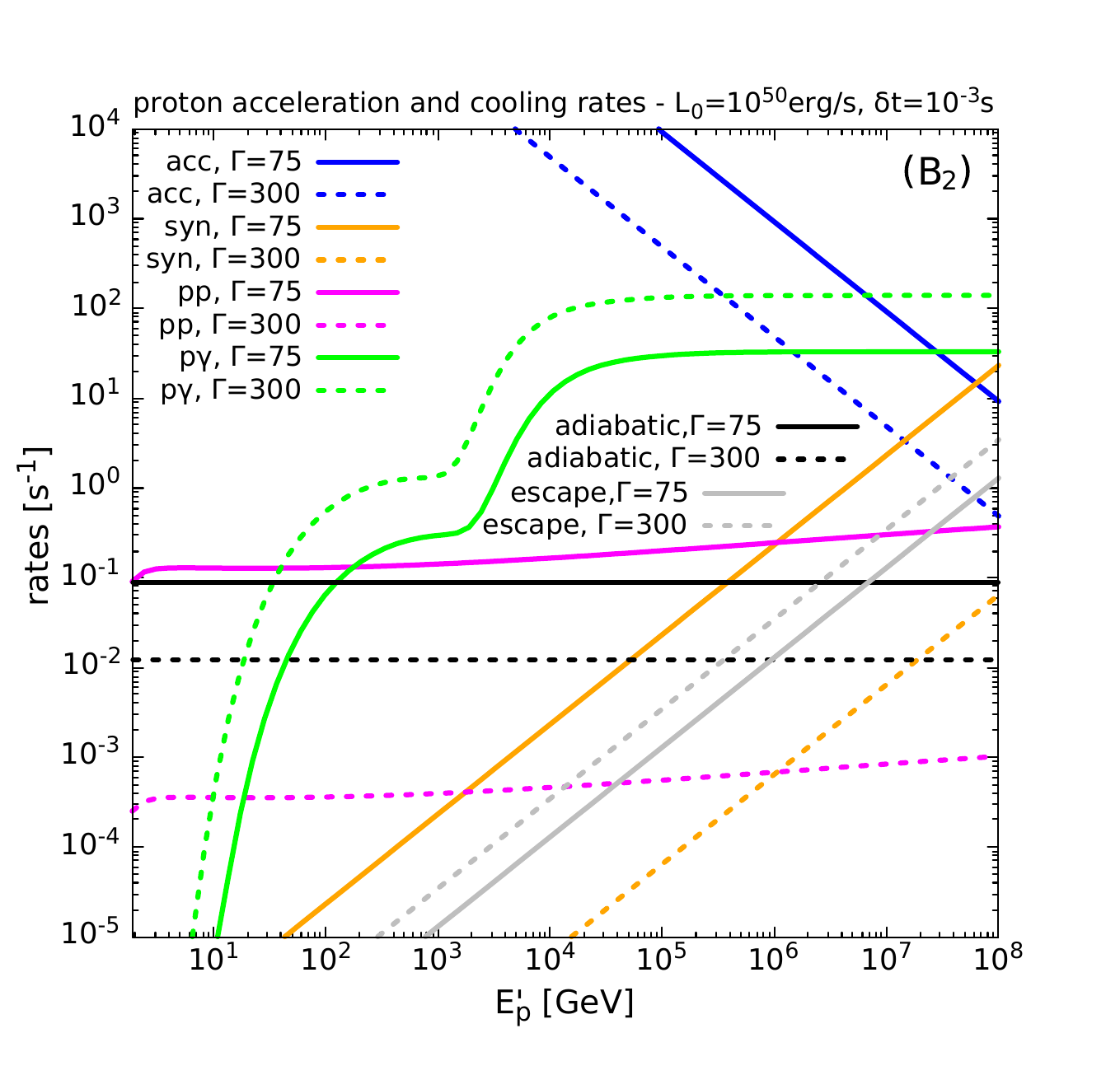} 
   \end{subfigure}     
 \caption{Cooling and acceleration rates for protons in a choked jet of a CCSN for the parameter sets $A_1$ and $A_2$ in the top-left and top-right panels, respectively. Bottom panels correspond to the sets $B_1$ and $B_2$ on the left and right, respectively. In each panel, solid and dashed curves refer to low and high values of $\Gamma$, respectively. Blue curves mark the acceleration rate, green ones correspond to the $p\gamma$ process, orange ones to synchrotron cooling, magenta curves refer to $pp$ interactions, and black ones to adiabatic cooling. The escape rates are marked by gray lines. }\label{fig1:p-rates}
\end{figure*}

According to the usually assumed scenario, a population of electrons, is accelerated in the shocked jet head and deposit all their energy there through synchrotron and inverse Compton (IC) radiation. This emission thermalizes since the optical depth is very high \citep[e.g.,][]{he2018}, 
\be
\tau_{\rm h}&\approx& (4\Gamma_{\rm rel}+3)n_{\rm j}(r_{\rm h}) \, \sigma_T  \nonumber \\ 
&\simeq&  294 \,  \Gamma_2^{-2}L_{0,50}^{3/5}\rho_{\rm ext,-7}^{2/5}r_{\rm ext,13.5}^{5/2}t_{\rm j,3}^{3/5},
\ee
where $\Gamma_{\rm rel}\approx \Gamma/(2\Gamma_{\rm h})$ is the Lorentz factor of the jet head with respect to the jet.
If a fraction $\epsilon_e\approx 0.1$ of the kinetic energy of the flow in the jet head is carried by the accelerated electrons, then it follows that the temperature can be expressed as \citep[e.g.,][]{meszaros2001,he2018}: 
\begin{equation}
T_{\rm h}= \left[\epsilon_e(\Gamma_{\rm rel}+ 3)(\Gamma_{\rm rel}-1) \left(\frac{L_{0}}{a\,\pi \Gamma^2 r_{\rm h}^2  c}\right)\right]^\frac{1}{4},
\end{equation}
which implies $k_{\rm B}T_{\rm h}\sim 70\,{\rm eV}$ for $L_0=10^{49}{\rm erg/s}$ and $k_{\rm B}T_{\rm h}\sim 190\,{\rm eV}$ for $L_0=10^{50}{\rm erg/s}$. 
This temperature characterizes the black body distribution of photons in the jet head frame, and we assume that a fraction $f_{\rm esc}=1/\tau_{\rm h}$ of them escape to the IS region, where protons are accelerated. Therefore, the distribution of photons in this region $(n_{\rm ph})$ is obtained using Lorentz invariants as explained in Ref. \citep{reynoso2023}, and the corresponding cooling rate due to $p\gamma$ interactions can then be computed as
\citep{atoyan2003}:
\begin{multline}
t_{p\gamma}^{-1} (E'_p) = \int_{E_{\rm th}/2\gamma_p}^{\infty} dE_{\rm ph} \frac{c\, n_{\rm ph}(E_{\rm ph})}{2 \gamma_p^2 E_{\rm ph}^2} \\ \times \int_{E_{\rm th}}^{2 E_{\rm ph} \gamma_p} dE_{\rm r} \sigma_{p\gamma}(E_{\rm r}) K_{p\gamma}(E_{\rm r}) E_{\rm r}. \label{tpg}
\end{multline}
Here, $\gamma_p=E'_p/(m_pc^2)$, and the threshold energy $E_{\rm th}=2m_ec^2$ corresponds to $e^{+}e$-pairs (Bethe-Heitler process) or to pion production with $E_{\rm th}\simeq 150{\,\rm MeV}$. The relevant cross sections $\sigma_{p\gamma}$ and inelasticity coefficients $K_{p\gamma}$ are taken following \cite{begelman1990}.

In this work, we assume that protons are accelerated by the IS at a rate 
\be
t_{\rm acc}^{-1}(E'_p)=\eta\,\frac{  e\, B' \,c}{E'_p},\label{tacc}
\ee 
where the efficiency coefficient is $\eta= 0.1$. These protons are supposed to carry a power $L_p=\epsilon_{\rm rel} L_{\rm j}$, with a fraction $\epsilon_{\rm rel}\approx 0.1$ of the total kinetic power carried by the jet. For simplicity, we neglect the role of accelerated electrons, which was studied in Ref. \citep{reynoso2023}\footnote{If the power injected in accelerated electrons is similar to the corresponding to protons, then the synchrotron photons emitted by electrons are additional targets for $p\gamma$ interactions. But this effect is not relevant for an electron contribution $\lesssim L_p/100$, similar to the inferred value from cosmic ray observations.}.    
We assume representative sets of parameters with the values shown in Table \ref{table:params}, in line with the normally adopted ones \citep{fasano2021,guetta2023}. Specifically, setting an energy budget of $E_{0}=10^{53}{\rm erg}$ in such a way that $E_0=L_0t_{\rm j}$, we consider two possibilities, $L_0=10^{49}{\rm erg/s}$ and $L_0=10^{50}{\rm erg/s}$, where the jet is active during $t_{\rm j}=10^3{\rm s}$ and $t_{\rm j}=10^4{\rm s}$, respectively. It can be checked that such situations actually correspond to choked jet, since $t_{\rm j}$ is shorter than the jet break out time \citep{murase2013},
\be
t_{\rm bo}\simeq 21100\,{\rm s} \, L_{0,50}^{-1/3}\left(\frac{\theta_{\rm op}}{0.2}\right)^{2/3}\left( \frac{R_*}{10^{13.5}{\rm cm} }\right), 
\ee
where $R_*$ is the radius of the red supergiant progenitor. 

Particle acceleration can be effective if IS are not radiation mediated \citep{levinson2008}. Given that in the collimated jet the flow is expected to be radiation dominated \citep{murase2013}, we require that $r_{\rm is}  < r_{\rm cs}$ and this leads to a maximum value of the jet Lorentz factor:
\be
 \Gamma\lesssim \Gamma_{\rm max}116 \, t_{\rm j,3}^{1/5}L_{0,50}^{3/20}\delta t_{-2}^{-1/2}.
\ee
Additionally, a low optical depth at the inner shocks' position is usually assumed to avoid radiation mediated shocks: 
\be 
  \tau_{\rm is}= \frac{r_{\rm is}}{\Gamma}n'_{\rm j}\sigma_{\rm T}< 1,
\ee
and this leads to a minimum value for the Lorentz factor, i.e.,
\be
\Gamma \gtrsim \Gamma_{\rm min} 46 \,L_{0,50}^{1/5}\delta t_{-2}^{-1/5}.  
\ee

Taking into account these considerations, we present in Table \ref{table:params} the different sets of parameters adopted:
Sets $A_{1,2}$ correspond to $\delta t= 10^{-2}{\rm s}$ with $L_0=10^{49}{\rm erg/s}$ and $L_0=10^{50}{\rm erg/s}$, and sets $B_{1,2}$ are analogous, but for $\delta t= 10^{-3}{\rm s}$. In particular, the possible values for the Lorentz factor $\Gamma$ are considered to lie in the range $(\Gamma_{\rm min}..\Gamma_{\rm max})$, as quoted in Table \ref{table:params} for the different parameter sets.

The injection of primary particles in the comoving frame is obtained as
\be 
Q'_i(E'_i)= \frac{d\mathcal{N}_i}{dE'_i d\Omega'\,dV'dt'}=K_iE'_i \exp\left(-\frac{E'_i}{E'_{i,\rm max}}\right),
\ee
where $E'_{i,\rm max}$ is a maximum energy determined by the balance of the acceleration rate with the total rate of escape plus energy loss. $K_i$ is a normalization constant that is fixed by the following expression used to obtain of the power carried by the relativistic particles, $L_i$: 
\be
\Delta V \int_{4\pi}d\Omega \int_{E_{i,\rm min}}^{\infty}dE_i E_iQ_i(E_i)= L_i. 
\ee
Here, $\Delta V= 4\pi r_{\rm is}^2 \Delta r$ is the volume of the region with IS, and 
$\Delta r\simeq c \delta t $ is the corresponding thickness in the central source (CS) rest frame. The injection can be transformed to the latter frame taking into account that 
\be
 \frac{E_i}{p^2}\frac{d\mathcal{N}_i}{dV\,dp\,d\Omega\,dt}\propto 
  \frac{Q_i}{\sqrt{E_i^2-m_i^2c^4}} \label{Qtransf}
\ee
is a Lorentz invariant \citep{dermer2002}, and that the energy in the comoving frame is $E'_i=\Gamma(E_i-\beta\mu \sqrt{E_i^2-m_i^2})$, where $\mu= \cos{\theta_i}$ is the cosine of the angle between the particle momentum and the jet velocity in the CS frame. The minimum energy of the accelerated particles is set as  $E'_{i,\rm min}=2m_i c^2$ in the comoving frame, as diffusive shock acceleration is effective only for suprathermal particles \citep[e.g.][]{kang1995}.  

In order to obtain the particle distributions, we solve the steady-state transport equation
\be
\frac{d\left[b'_{i,\rm loss}(E'_i) N'_i(E'_i)\right]}{dE'_i}+  \frac{N'_i(E'_i)}{T_{i,\rm esc}}= Q'_i(E'_i),\label{Eq_transport_i}  
\ee
where the $b'_{i,\rm loss}$ is the total energy loss for particles of the type $i$, i.e., 
\be 
b'_{i,\rm loss}= -\left.\frac{dE'_i}{dt'}\right|_{\rm loss}\equiv  E'_i \sum_j t'^{-1}_{i,j}(E'_{i}).
\ee
Here, the cooling rates $t'^{-1}_{i,j}$ correspond to the different processes $j$.
The escape timescale is  considered to be the Bohm diffusion time corresponding to the size $\Delta r'$, and the associated escape rate is generally negligible as compared to the cooling rates involved. Apart from the above mentioned $p\gamma$ process, the rates for other possible processes are taken as
\be
t_{i,{\rm syn}}^{-1}(E'_i)= \frac{4}{3} \left(\frac{m_e}{m_i}\right)^3 \frac{\sigma_T B^2}{m_e c \ 8\pi} \frac{E'_i}{m_i c^2} 
\ee
for synchrotron cooling,
\be
t_{pp}^{-1}(E'_p)\simeq \frac{1}{2} n'_{\rm j} c\,\sigma_{pp}(E'_p)
\ee
for cooling through $pp$ interactions, 
and 
\be
t_{\rm ad}^{-1}\approx \frac{c}{r_{\rm is}}
\ee
for adiabatic cooling.

We show in Fig. \ref{fig1:p-rates} the acceleration and cooling rates for protons adopting the sets of parameters described in Table \ref{table:params}, specifying two possible values for the Lorentz factor $\Gamma$ within each set. That is, for the sets $A_1$ and $A_2$, which appear in the top panels, we adopted $\Gamma=\{ 50,125\}$ and $\Gamma=\{ 50,115\}$, respectively. It can be seen that $p\gamma$ interactions give always the dominant cooling mechanism for high energy protons. The corresponding the maximum energy $E'_{p,{\rm max}}$ is found to be higher for the lowest values of $\Gamma$ adopted, since the magnetic field is higher in such cases and this yields a higher acceleration rate (see Eqs. \ref{Bmag},\ref{tacc}).

\begin{figure*}[]                            
\centering
\  \centering
    \begin{subfigure}[t]{0.49\textwidth}
        \centering                          
        \includegraphics[width=0.5\linewidth,trim= 180 30 180 35]{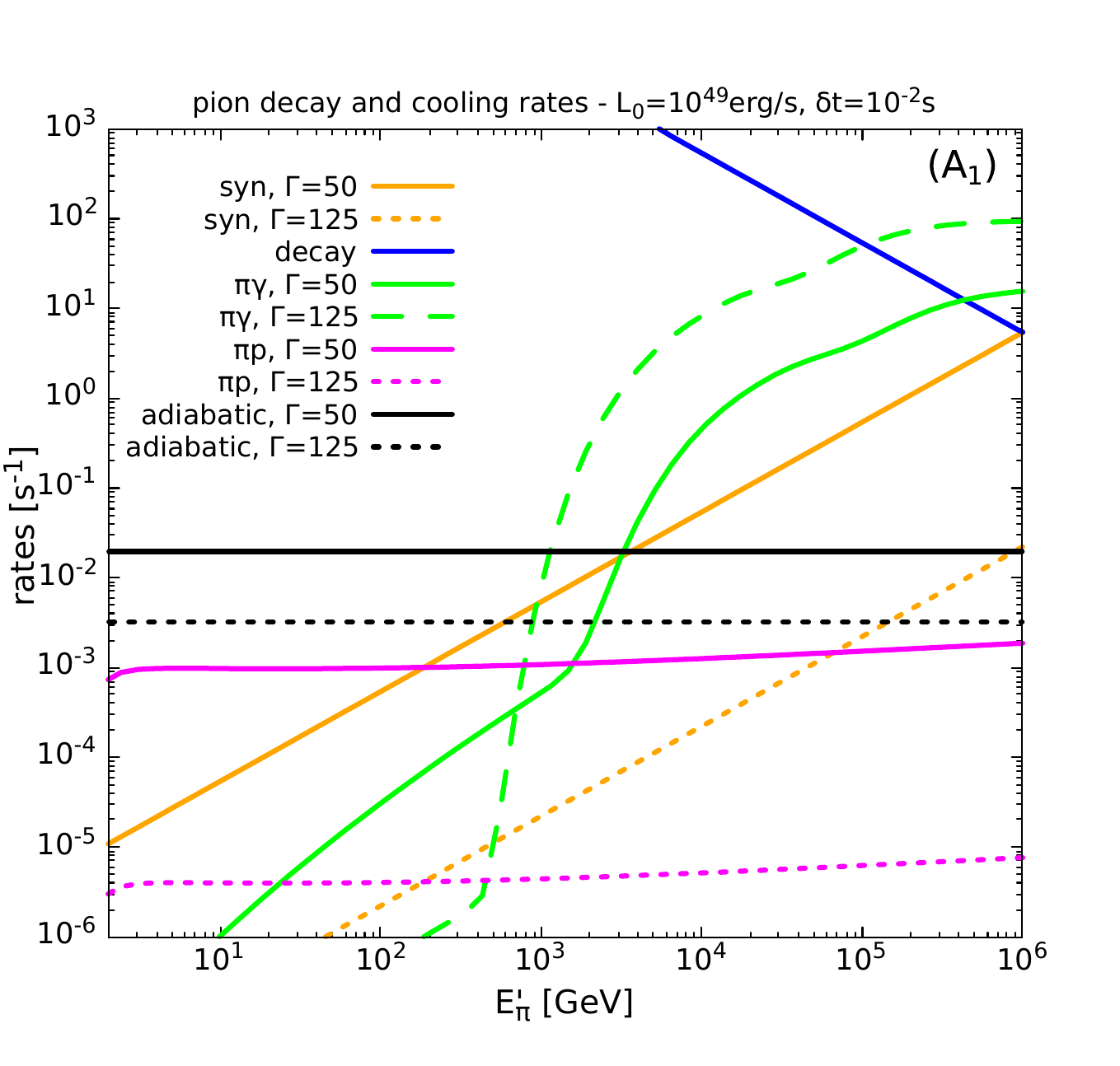} 
    \end{subfigure}
    \hfill
    \begin{subfigure}[t]{0.49\textwidth}
        \centering
        \includegraphics[width=0.5\linewidth,trim= 180 30 180 35]{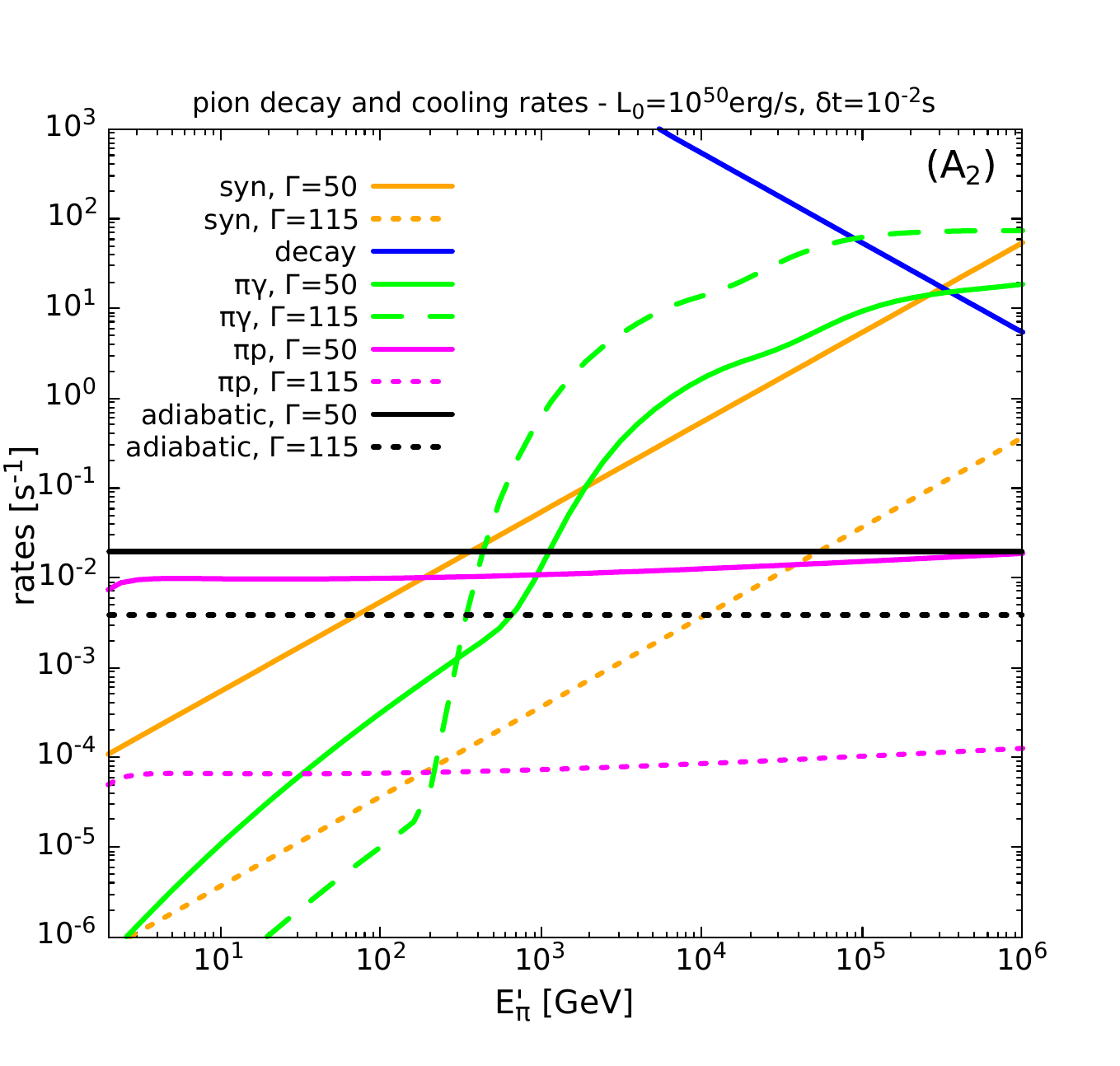} 
    \end{subfigure}
    \begin{subfigure}[t]{0.49\textwidth}
        \centering                     
        \includegraphics[width=0.5\linewidth,trim= 180 30 180 35]{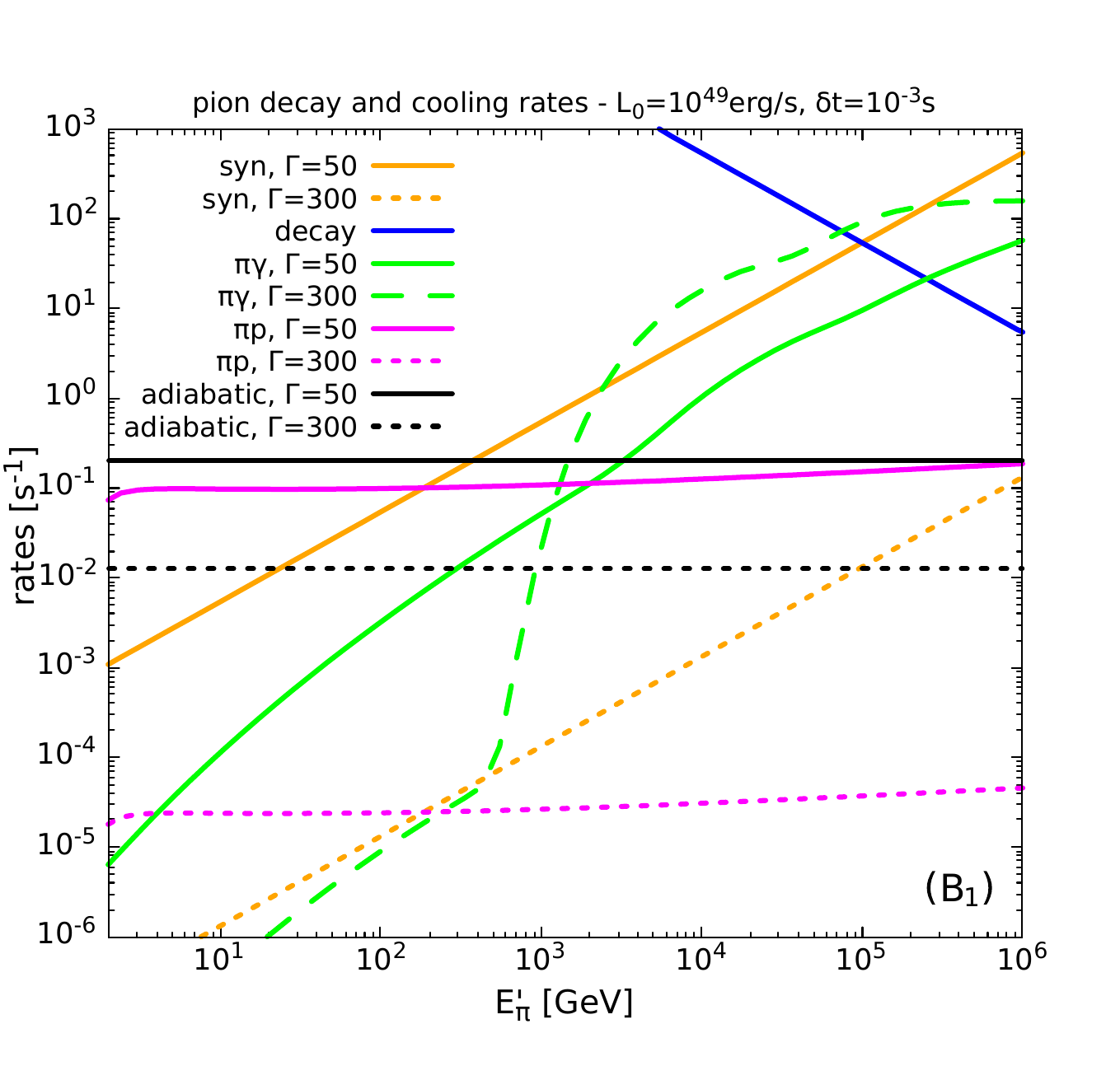} 
    \end{subfigure}
    \hfill
    \begin{subfigure}[t]{0.49\textwidth}
        \centering
        \includegraphics[width=0.5\linewidth,trim= 180 30 180 30]{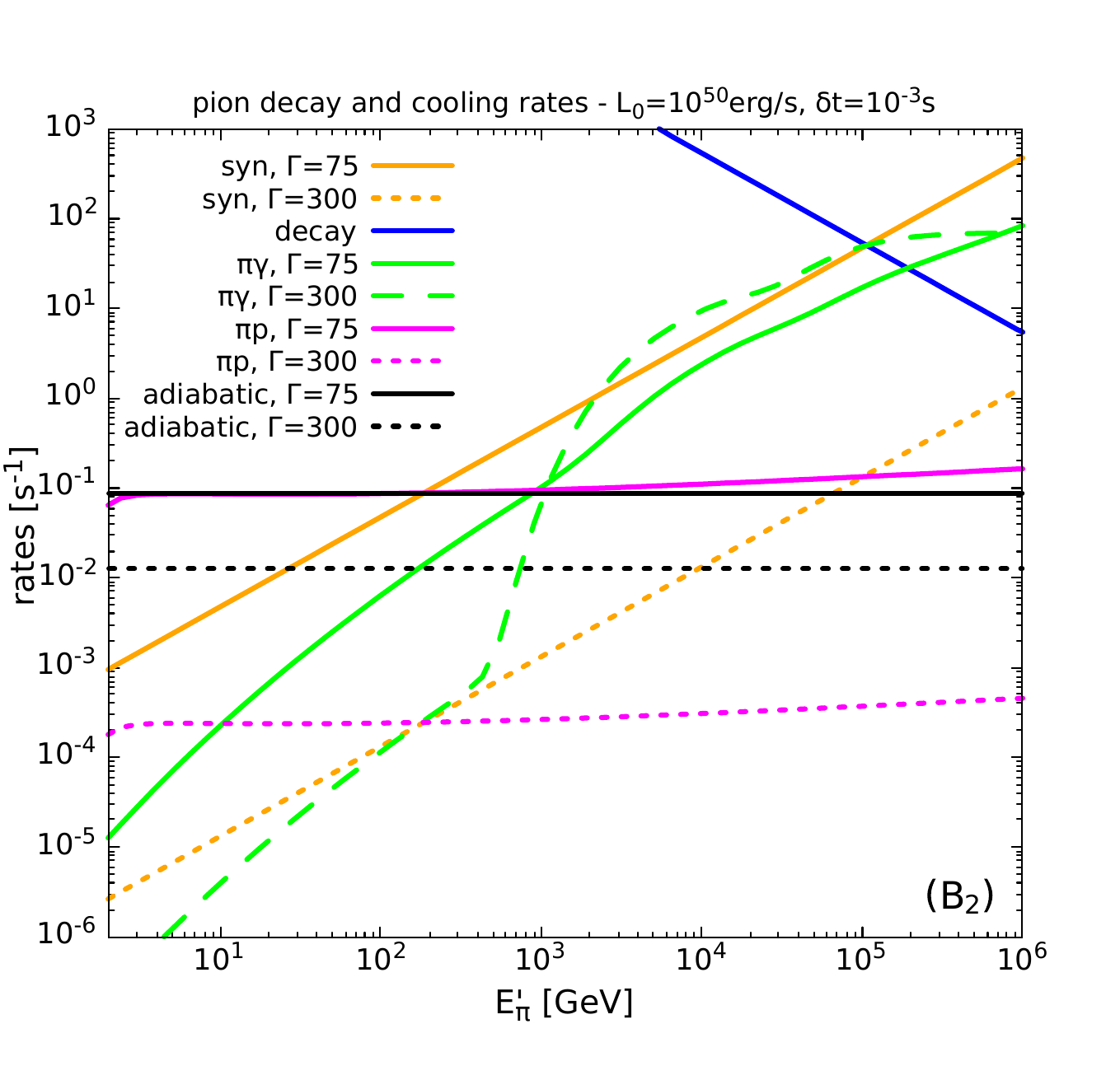} 
   \end{subfigure}     
 \caption{Cooling and decay rates for pions in a choked jet of a CCSN for the parameter sets the parameter sets $A_1$ and $A_2$ in the top-left and top-right panels, respectively. Bottom panels correspond to the sets $B_1$ and $B_2$ on the left and right, respectively. In each panel, solid and dashed curves refer to low and high values of $\Gamma$, respectively. Blue curves mark the decay rate, green ones correspond to the $\pi\gamma$ process, orange ones to synchrotron cooling, magenta curves refer to $\pi p$ interactions, and black ones to adiabatic cooling.}\label{fig2:pi-rates}
\end{figure*}
\begin{figure*}[]                            
\centering
\  \centering
    \begin{subfigure}[t]{0.49\textwidth}
        \centering                          
        \includegraphics[width=0.5\linewidth,trim= 180 30 180 35]{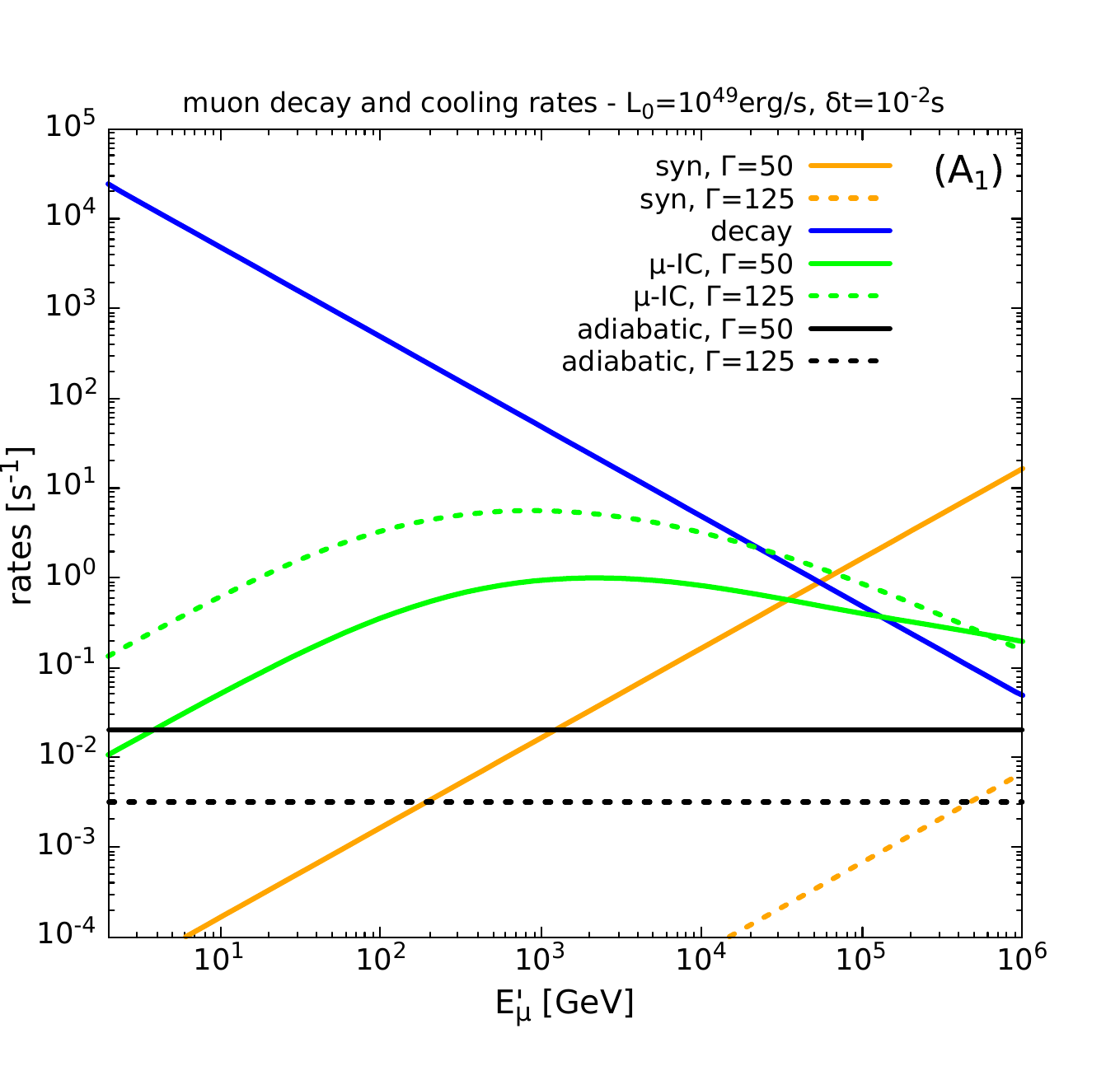} 
    \end{subfigure}
    \hfill
    \begin{subfigure}[t]{0.49\textwidth}
        \centering
        \includegraphics[width=0.5\linewidth,trim= 180 30 180 35]{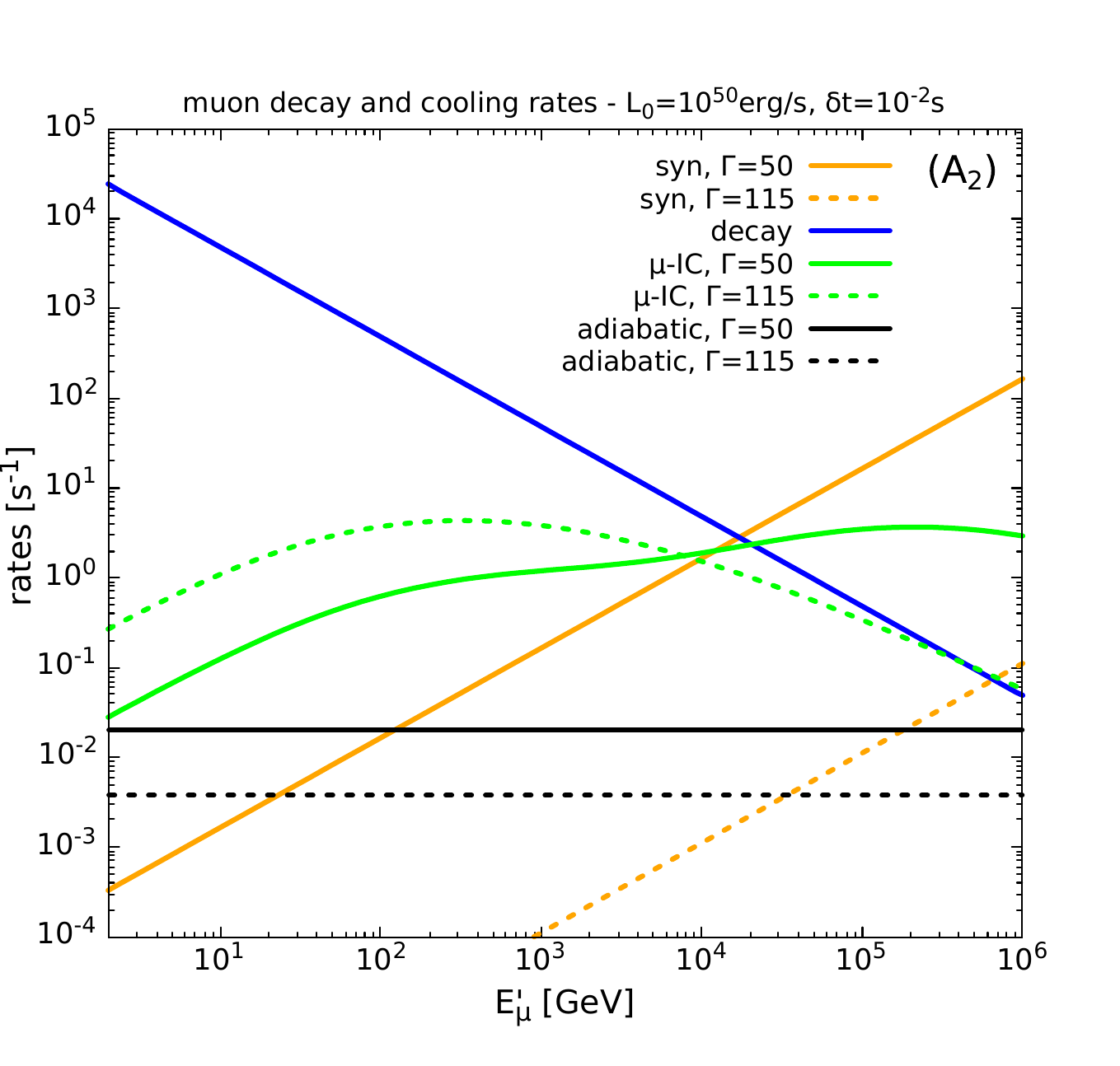} 
    \end{subfigure}
    \begin{subfigure}[t]{0.49\textwidth}
        \centering                     
        \includegraphics[width=0.5\linewidth,trim= 180 30 180 35]{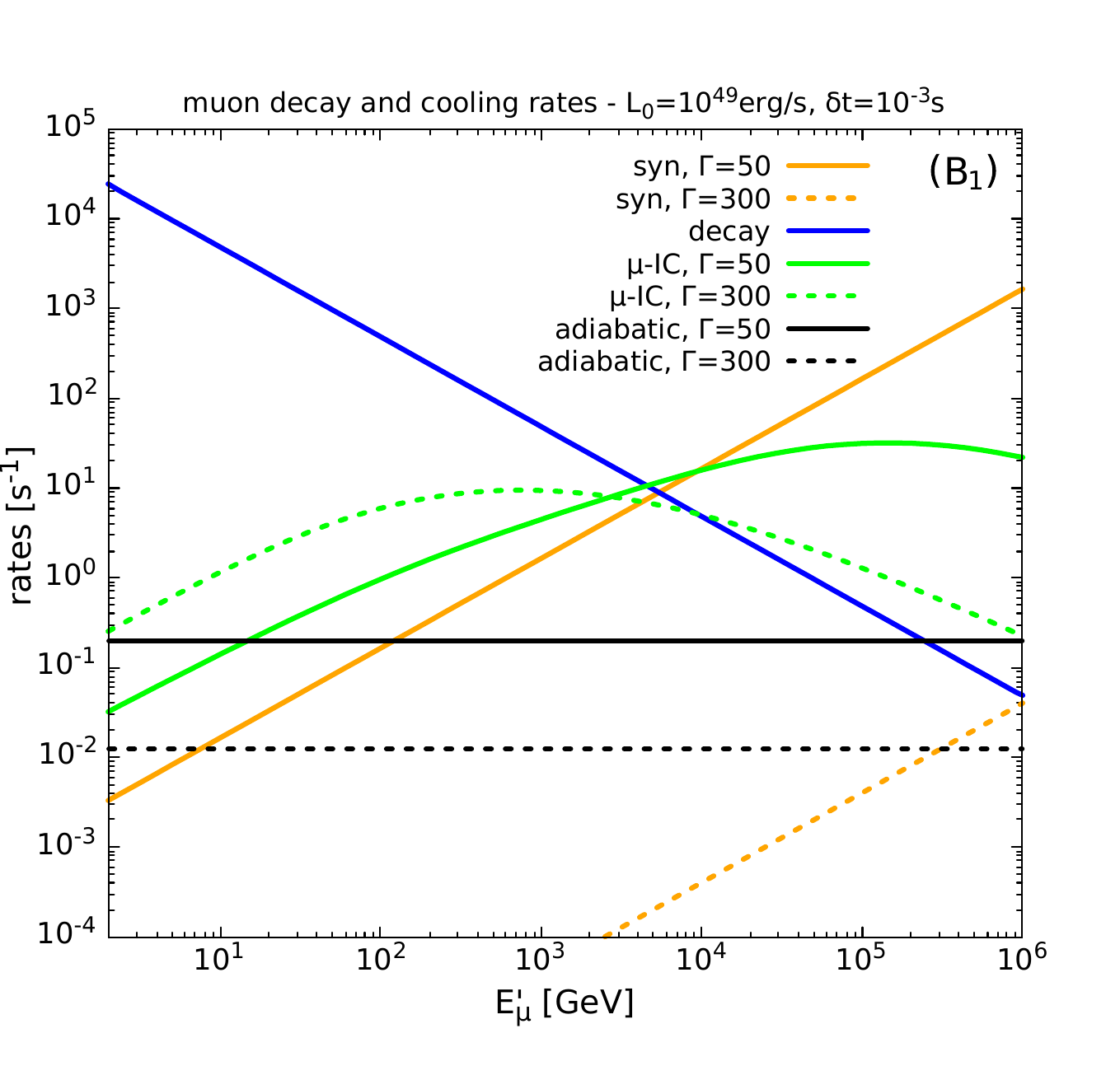} 
    \end{subfigure}
    \hfill
    \begin{subfigure}[t]{0.49\textwidth}
        \centering
        \includegraphics[width=0.5\linewidth,trim= 180 30 180 30]{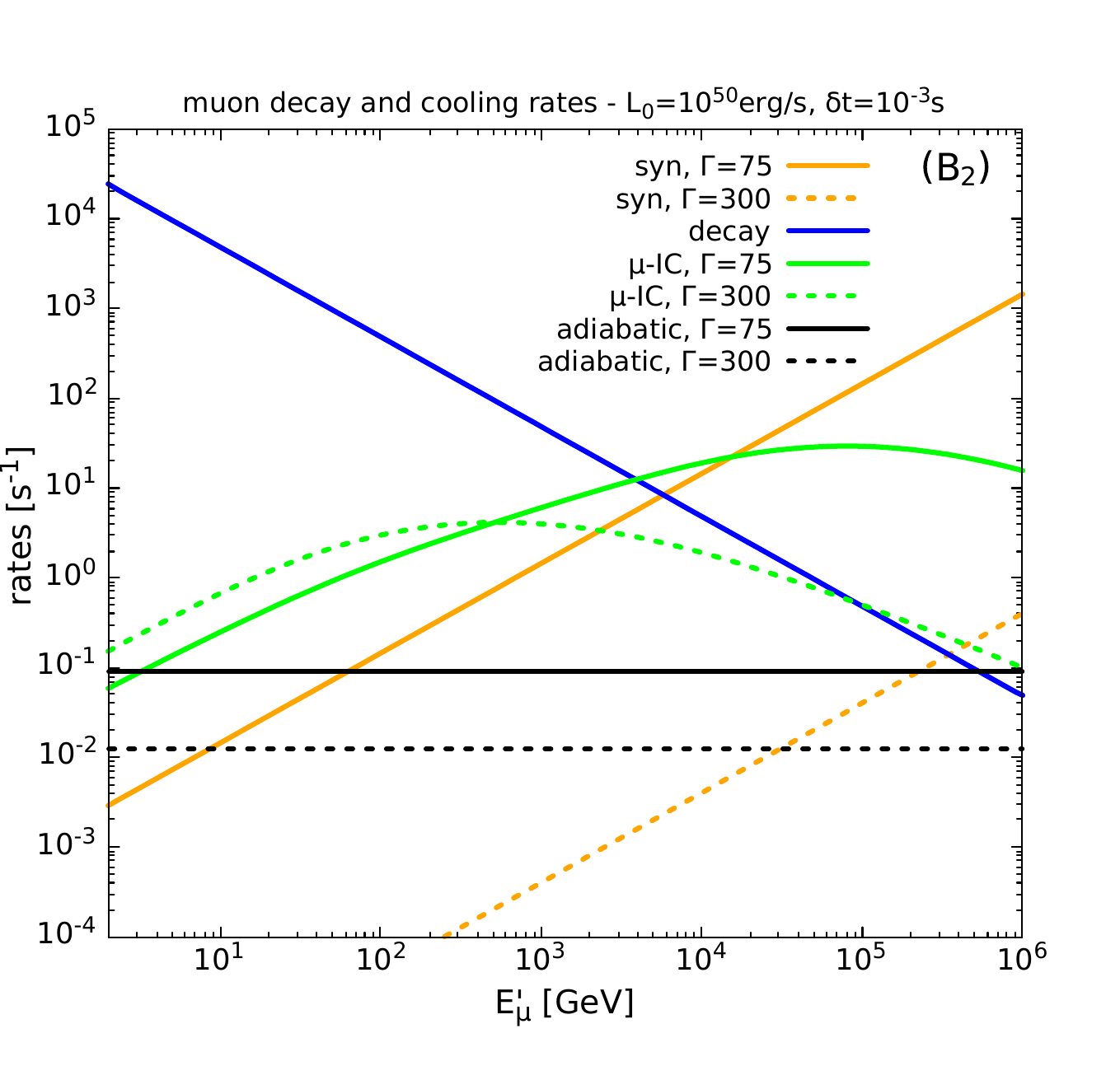} 
   \end{subfigure}     
 \caption{Cooling and decay rates for muons in a choked jet of a CCSN for the parameter sets $A_1$ and 
 $A_2$ in the top-left and top-right panels, respectively. Bottom panels correspond to the sets $B_1$ and $B_2$ on the left and right, respectively. In each panel, solid and dashed curves refer to low and high values of $\Gamma$, respectively. Blue curves mark the decay rate, green ones correspond to the $\mu$-IC process, orange ones to synchrotron cooling, and black ones to adiabatic cooling.}\label{fig3:mu-rates}
\end{figure*}


The solution of Eq.(\ref{Eq_transport_i}) can be expressed as
\begin{multline}
N'_i(E'_i)= \frac{1}{b_{i,\rm loss}(E'_i)}\int_{E'_i}^\infty dE'Q'_i(E')  \\
 \times \exp\left[-\int_{E'_i}^{E'}\frac{dE''}{T_{\rm esc}(E'')b_{i,\rm loss}(E'')}\right],\label{solution_Eq_transport}
\end{multline}
and we apply it to the different particle species $(e, \, p, \, \pi^{\pm}, \, \mu^{\pm})$ taking into account the corresponding cooling processes, injections and possible decays in each case.

In particular, the injection of pions due to $p\gamma$ interactions, $Q_{p\gamma\rightarrow \pi^{\pm}}$, is obtained by applying the expressions of Ref. \cite{hummer2010}, as described in Ref. \citep{reynoso2023}. The relevant cooling mechanisms for pions are shown in Fig. \ref{fig2:pi-rates}, where it can be seen that $\pi\gamma$ interactions are dominant at high energies, although for low values of $\Gamma$, synchrotron cooling can also become the leading process.

As for muons, the injections $Q'_{\pi^{\pm}(E'_\pi)\rightarrow\mu^{\pm}}$ again are obtained applying the formulae of Ref. \cite{lipari2007} as in Ref. \citep{reynoso2023}. In Fig. \ref{fig3:mu-rates}, we show the muon decay and cooling processes, where it can be seen that the leading loss mechanism is muon IC for low energies, where muon decay is faster. However, for higher energies, synchrotron cooling can also become dominant for the cases of lower $\Gamma$ values considered.
 
\section{Neutrino emission}\label{sec:nuflux}
\begin{figure}[!h]                            
\centering
\includegraphics[width=0.28\linewidth,trim= 180 100 180 50]{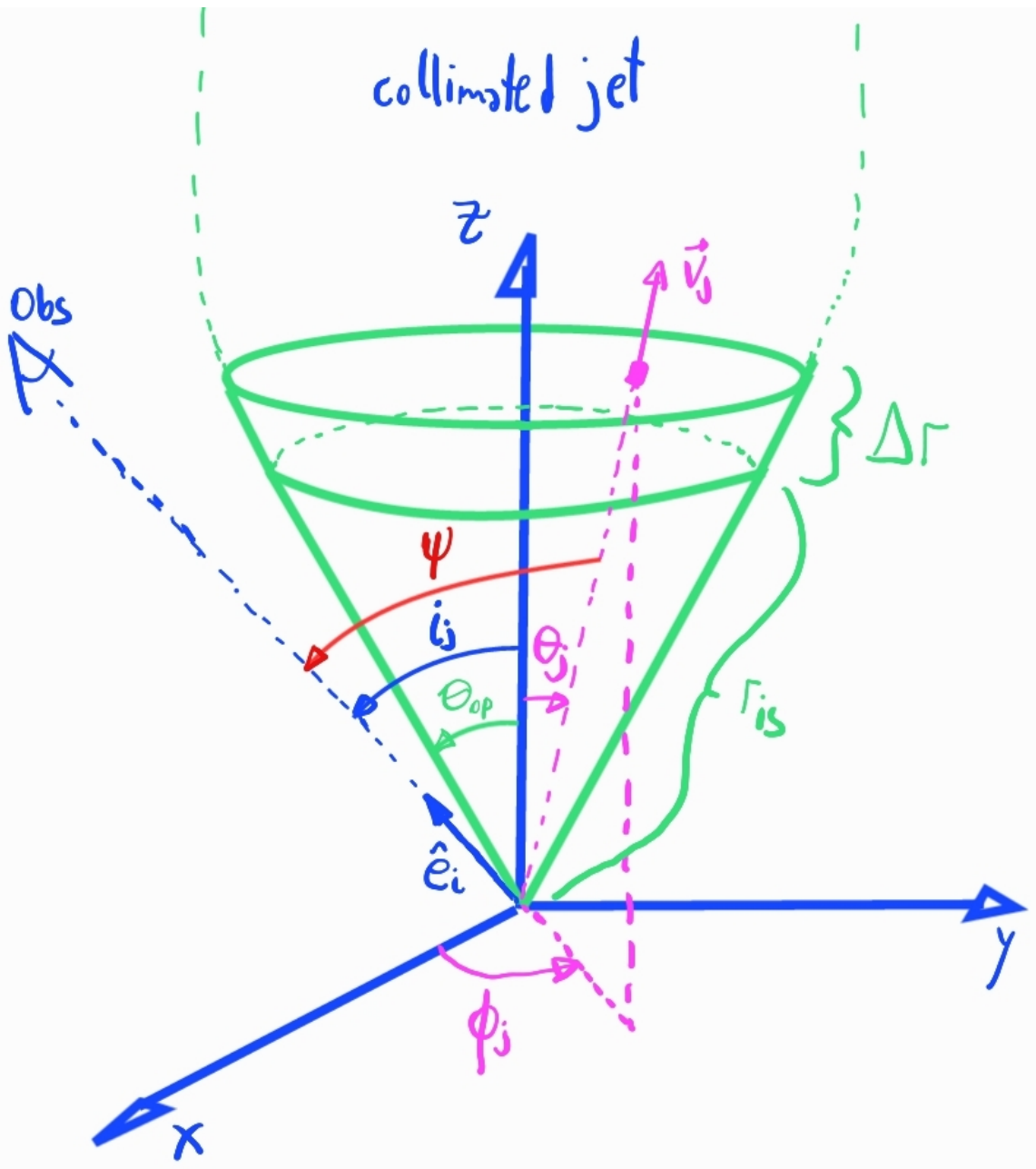} 
 \caption{Sketch of the emitting ejecta for an arbitrary line-of-sight direction at an angle $i_{\rm j}$ with respect to the jet axis. Not to scale. }\label{fig4:sketch-calc}
\end{figure}

In this section we describe the calculation of the neutrino flux to be observed from choked jets of CCSNe.
Once we have the pion and muon distributions $N'_{\pi^\pm}(E'_\pi)$ and $N'_{\mu^\pm}(E'_\mu)$ in the jet comoving frame, we can obtain the corresponding neutrino emissivities $Q'_{\pi\rightarrow \nu_\mu}(E'_\nu)$, $Q'_{\mu\rightarrow\nu_e}(E'_\nu)$, and $Q'_{\mu\rightarrow\nu_\mu}(E'_\nu)$ as described in Ref. \citep{reynoso2023}, following the expressions in Ref. \cite{lipari2007}, without distinguishing between neutrinos and anti-neutrinos.

We now consider in some detail the calculation of the  neutrino flux to be registered by an observer whose line of sight makes an angle $i_{\rm j}$ with the axis of the ejecta, such that a unit vector pointing to the observer reads
\be
\hat{e}_i=\sin i_{\rm j}\hat{e}_x+\cos i_{\rm j}\hat{e}_z.
\ee
A simple sketch is shown in Fig. \ref{fig4:sketch-calc}, where the IS in the jet take place at a spherical radius $r_{\rm is}$ within a half-opening angle $\theta_{\rm op}$. 
The usual assumption at this point is that
the bulk motion of the jet is directed radially, i.e. with a velocity  $\vec{v}_j=v_{\rm j}\hat{e}_r$, where 
$$\hat{e}_r=\sin\theta_{\rm j}\cos{\phi_{\rm j}} \hat{e}_x+ \sin\theta_{\rm j}\cos{\phi_{\rm j}} \hat{e}_y+\cos\theta_{\rm j}\hat{e}_z$$ is the unit vector of spherical coordinates. Therefore, the angle $\psi$ between the line of sight and the bulk motion depends on the position within the production volume:
\be 
\cos\psi= \hat{e}_i\cdot \hat{e}_r=\sin\theta_{\rm j}\cos{\phi_{\rm j}}\sin{i_{\rm j}}+\cos{\theta_{\rm j}}\cos{i_{\rm j}}.\label{cospsi}
\ee
Now, we can obtain the energy flux per unit frequency carried by neutrinos in the observer frame as
\be 
F_E(E_\nu)\equiv \frac{dE_\nu}{dA\,dt\,df}=\frac{1}{d_L^2}\int dV \,j_f(E_\nu(1+z)), 
\ee
where $z$ is the redshift of the source, $d_L$ is the luminosity distance, and $f=E_\nu/h$ is the neutrino frequency. The emissivity in the frequency $j_f= {dE_\nu}/({d\Omega\,dt\,df\,dV})$ is related to the energy emissivity used in this work as
\be
j_f(E)= h\,E\,Q_\nu(E).  
\ee
Making use of Eq. (\ref{Qtransf}) and neglecting the neutrino masses, we can express $F_E(E_\nu)$ in terms of the neutrino emissivity in the comoving frame, $Q'_\nu(E'_\nu)$, as
\be 
F_E(E_\nu)= \frac{1}{d_L^2}\int dV h\,E'_\nu D^2(\cos\psi)Q'_\nu\left(E'_\nu\right),\label{F_E}
\ee
where $E'_\nu={E_\nu(1+z)}/{D(\cos\psi)}$ is the comoving energy and the Doppler factor is 
\be 
 D(\cos\psi)=\frac{1}{\Gamma(1-\beta\cos{\psi})}.
\ee
We note that Eq. (\ref{F_E}) is consistent with Eqs. (1-2) of Ref. \cite{ahlers2019}.
In turn, $F_E(E_\nu)$ can be related to the differential neutrino flux $\varphi_\nu$ as 
\be 
E_\nu\varphi_\nu=E_\nu \frac{dN_\nu}{dE_\nu\, dA\,dt}=\frac{F_E}{h},
\ee 
and performing the radial part of the volume integral, we can write flux of muon neutrinos as
\begin{multline}
E_\nu \varphi_{\nu_\mu}(E_\nu,i_{\rm j})\simeq \frac{r_{\rm is}^2 \Delta r}{d_L^2} 
 \int_0^{2\pi}d\phi_j\int_0^{\theta_{\rm op}}d\theta_{\rm j}\sin{\theta_{\rm j}}\,E'_\nu D^2(\cos\psi) \\
 \times \left[Q'_{\nu_e}(E'_\nu)P_{\nu_e\rightarrow\nu_\mu}+ Q'_{\nu_\mu}(E'_\nu)P_{\nu_\mu\rightarrow\nu_\mu}\right],\label{nuflux}
\end{multline}
where the Doppler factor depends on the angles $\theta_{\rm j}$, $\phi_{\rm j}$, and the viewing angle $i_{\rm j}$ through Eq. (\ref{cospsi}).  We also include in Eq. (\ref{nuflux}) the effect of neutrino oscillations using the probabilities $P_{\nu_e\rightarrow\nu_\mu}\simeq 0.171 $ and $P_{\nu_\mu\rightarrow\nu_\mu}\simeq 0.453$, which arise using the best-fit values for the mixing angles given in Ref. \cite{esteban2020}.

We have carefully performed the integration of Eq. (\ref{nuflux}), specially in the cases where the line of sight is within the opening angle of the jet, i.e., for $i_{\rm j}<\theta_{\rm op}$ some part of the ejecta points exactly towards the observer. In these cases, the integrand is highly peaked for $\theta_{\rm j}\approx i_{\rm j}$ and $\phi_j\approx 0$.

Taking into account the jet duration $t_{\rm j}$ in the cases considered, we can compute the energy fluence or time-integrated flux as 
\be 
\mathcal{F}_{\nu_\mu}(E_\nu,i_{\rm j})=t_{\rm j}(1+z)E_\nu^2\varphi_{\nu_\mu}(E_\nu,i_{\rm j}) \, (\rm GeV\, cm^{-2}).
\ee 
This is useful, for instance, in order to compare with the upper limit given by IceCube for SN 2023ixf, 
\be 
 \left.\mathcal{F}_{\nu_\mu}\right|_{\rm UL}= 7.3\times 10^{-2}{\rm GeV\, cm^{-2} }, 
\ee
which is valid for $600\,{\rm GeV}<E_\nu<250\,{\rm TeV}$. Since this upper limit was obtained assuming a strict $E^{-2}$ dependence of the neutrino flux, it can be useful only as a reference in our cases.

\begin{figure*}[!h]                            
\centering
\  \centering
    \begin{subfigure}[t]{0.49\textwidth}
        \centering            \includegraphics[width=0.5\linewidth,trim= 180 30 192 35]{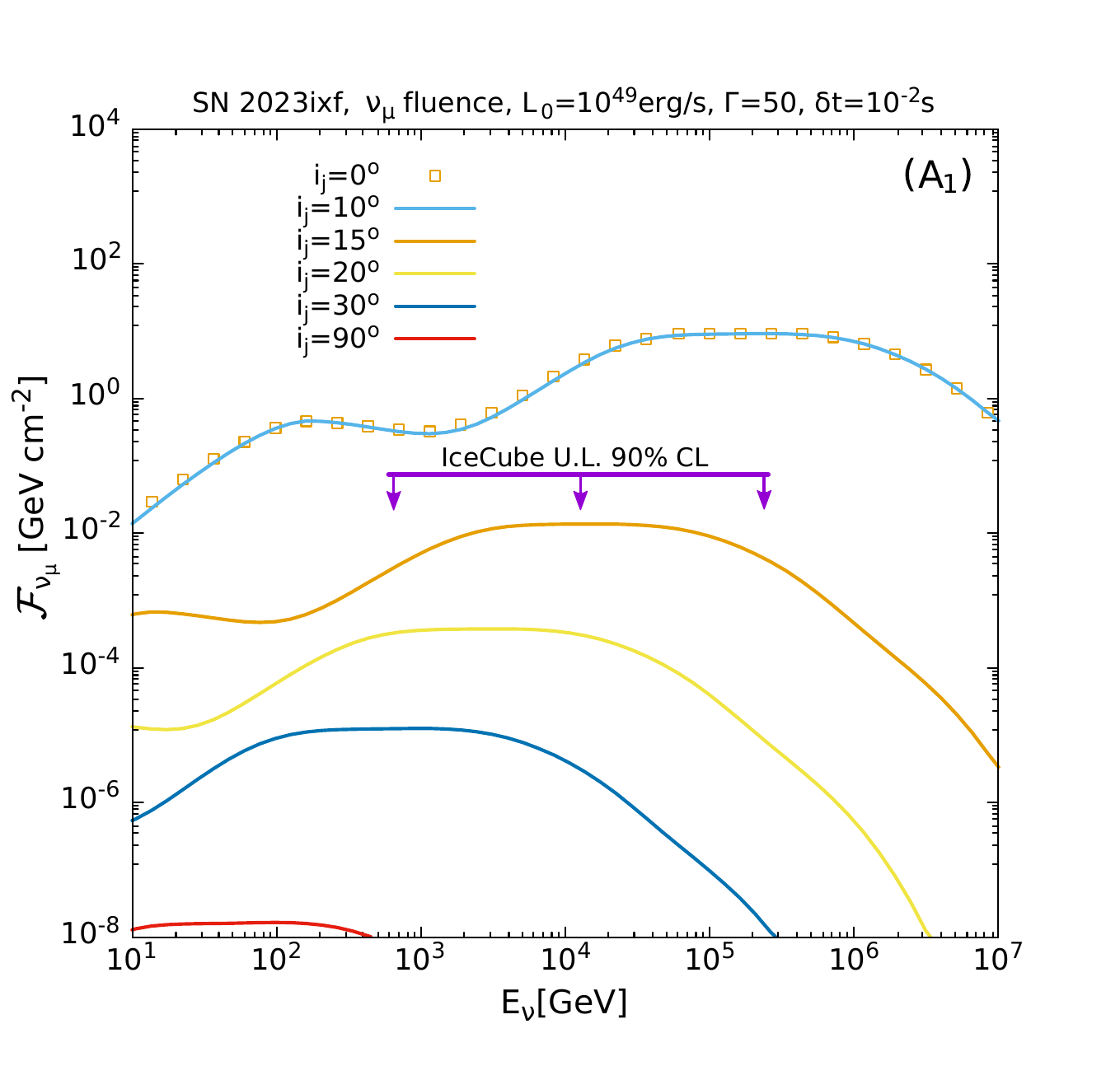} 
    \end{subfigure}
    \hfill
    \begin{subfigure}[t]{0.49\textwidth}
        \centering
    \includegraphics[width=0.5\linewidth,trim= 180 30 192 35]{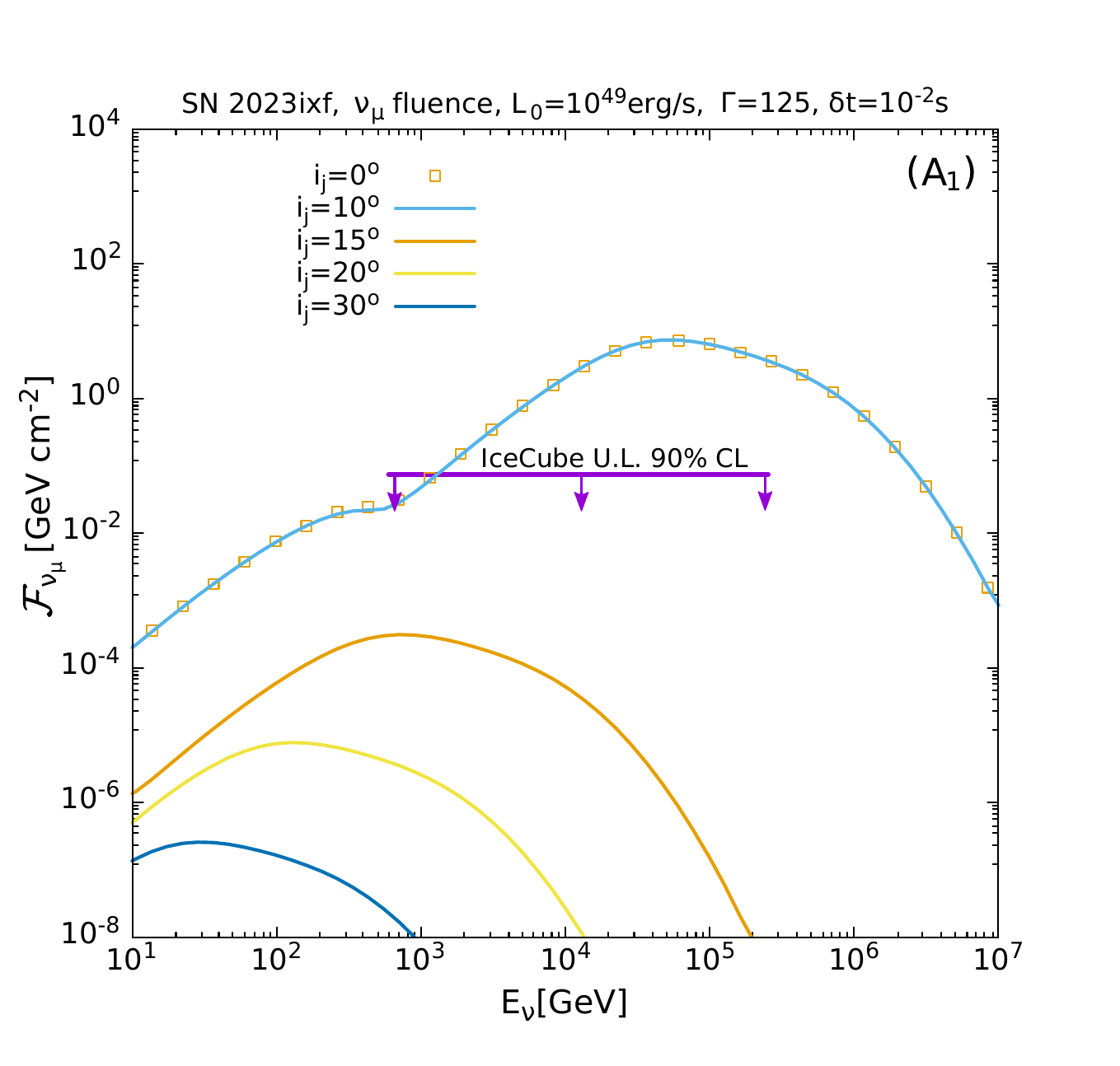} 
    \end{subfigure}
    \begin{subfigure}[t]{0.49\textwidth}
        \centering            \includegraphics[width=0.5\linewidth,trim= 180 30 192 35]{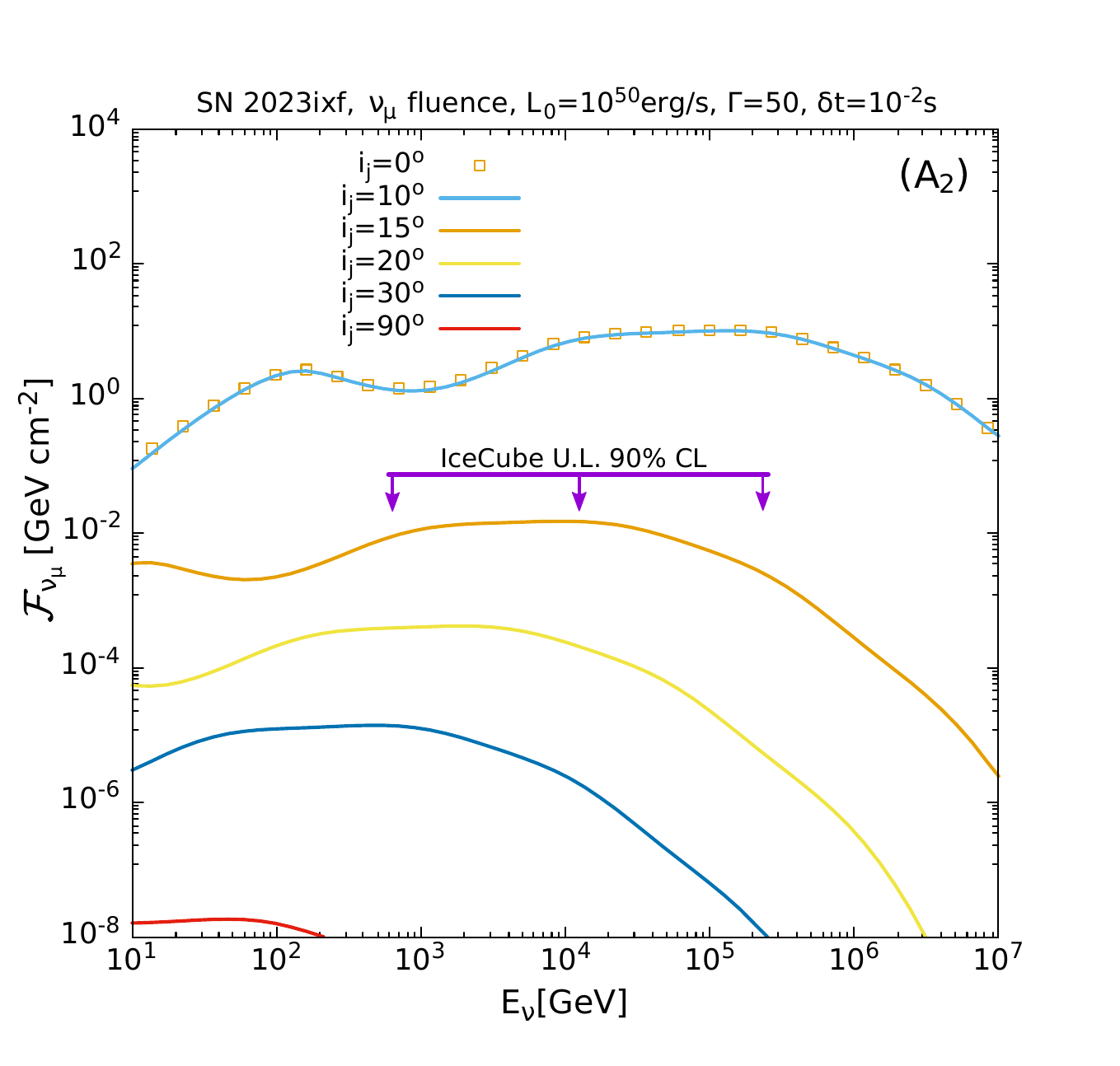} 
    \end{subfigure}
    \hfill
    \begin{subfigure}[t]{0.49\textwidth}
        \centering
        \includegraphics[width=0.5\linewidth,trim= 180 30 192 30]{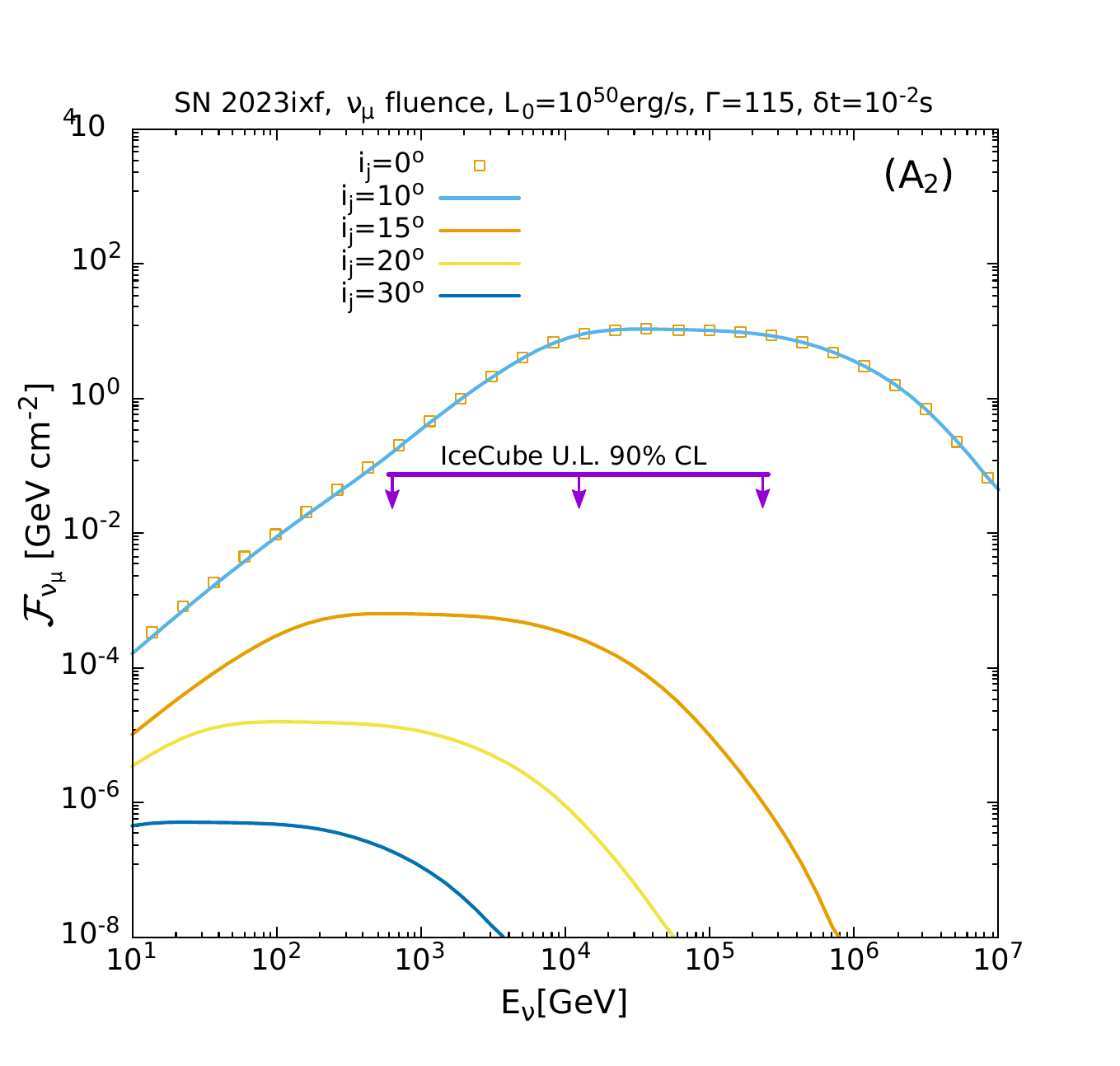} 
   \end{subfigure}     
 \caption{Different outcomes of the time-integrated flux of muon neutrinos for the sets of parameters $A_1$ and $A_2$ in the top and bottom panels, respectively. In the left panels, $\Gamma=50$, while we take $\Gamma=125$ in the top-right panel and $\Gamma=115$ in the bottom-right one.}\label{fig5:nufluence_A1A2}
\end{figure*}

\begin{figure*}[!h]                            
\centering
\  \centering
    \begin{subfigure}[t]{0.49\textwidth}
        \centering            \includegraphics[width=0.5\linewidth,trim= 180 30 192 35]{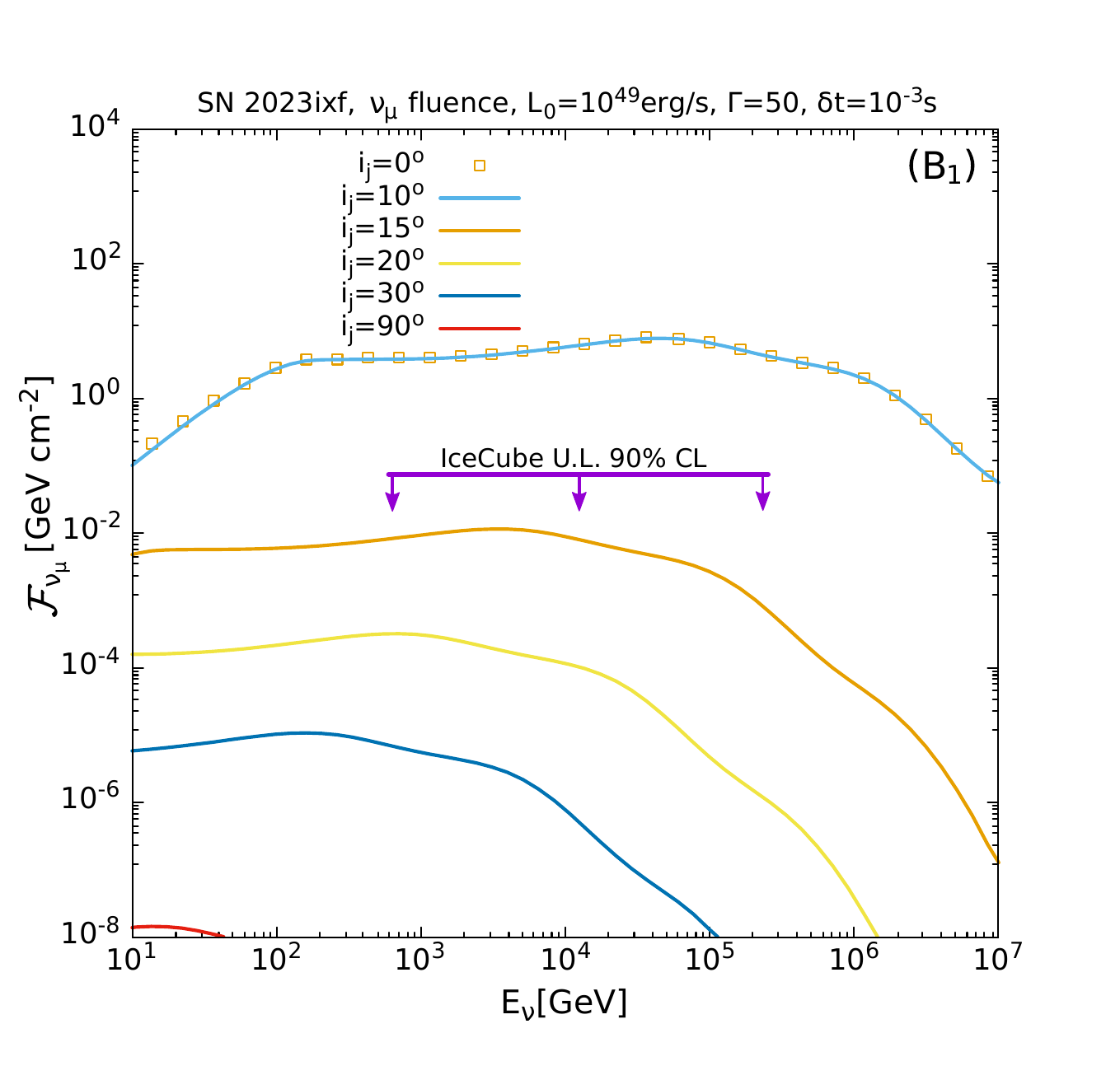} 
    \end{subfigure}
    \hfill
    \begin{subfigure}[t]{0.49\textwidth}
        \centering
    \includegraphics[width=0.5\linewidth,trim= 180 30 192 35]{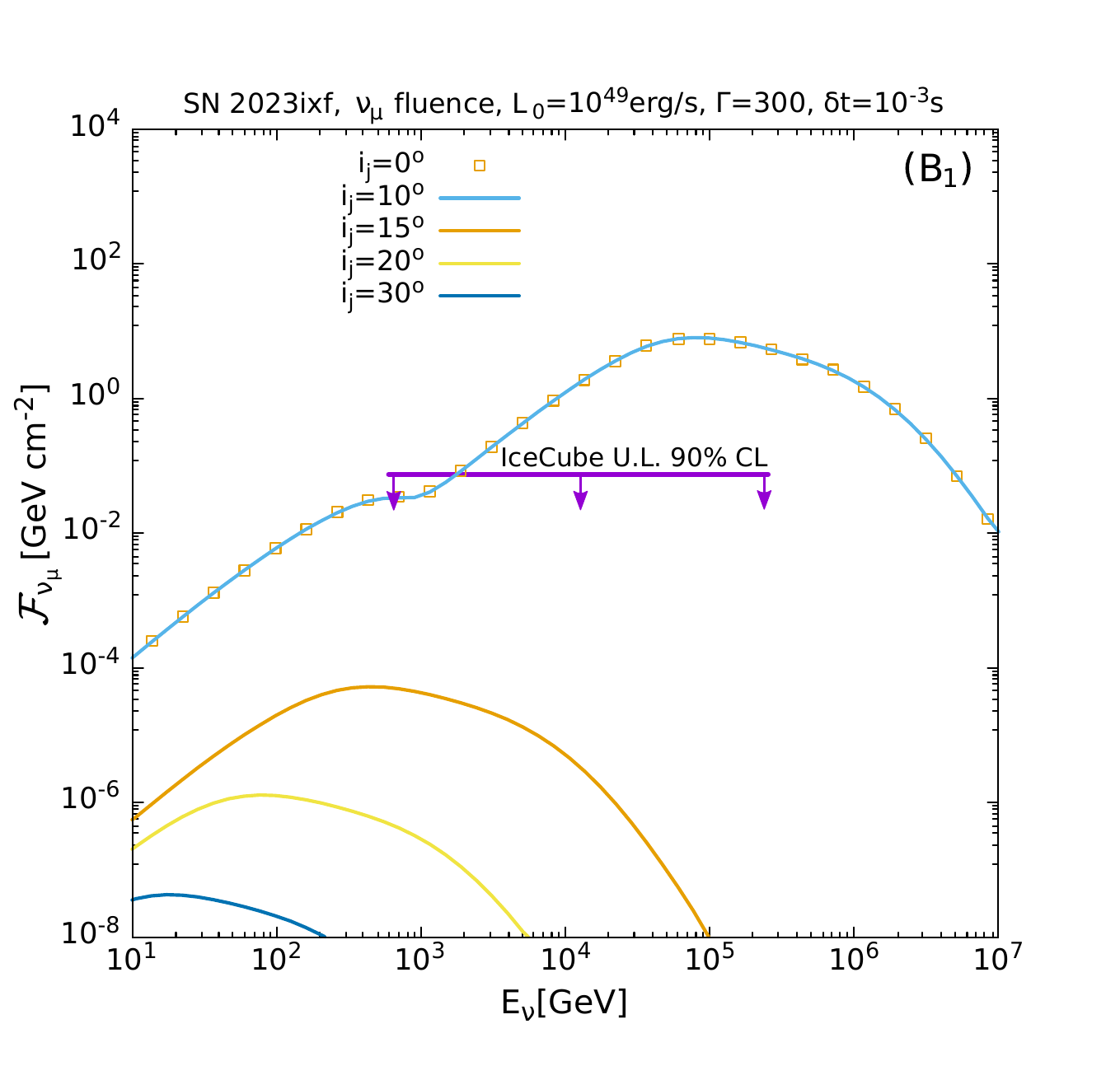} 
    \end{subfigure}
    \begin{subfigure}[t]{0.49\textwidth}
        \centering            \includegraphics[width=0.5\linewidth,trim= 180 30 192 35]{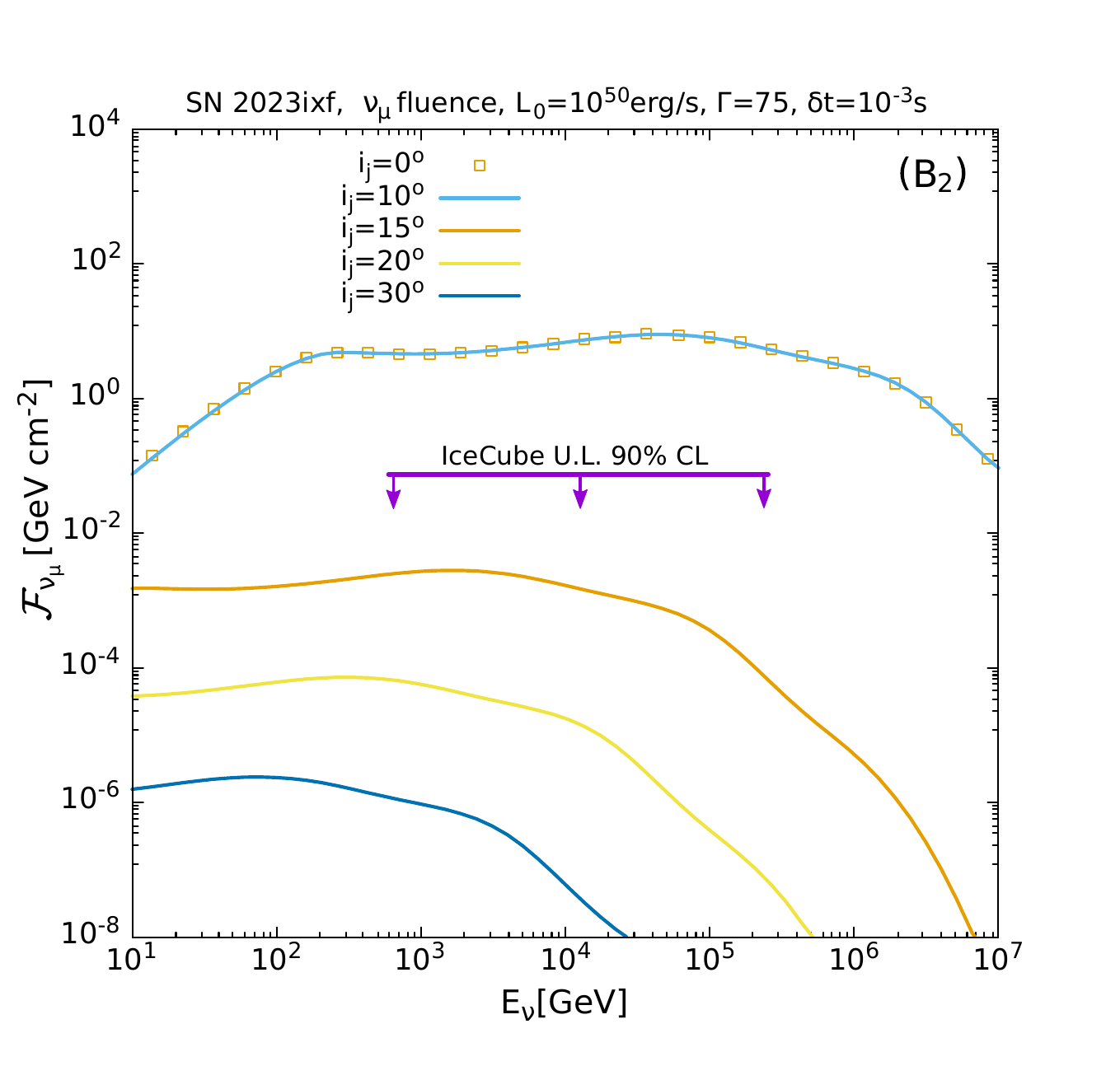} 
    \end{subfigure}
    \hfill
    \begin{subfigure}[t]{0.49\textwidth}
        \centering
        \includegraphics[width=0.5\linewidth,trim= 180 30 192 30]{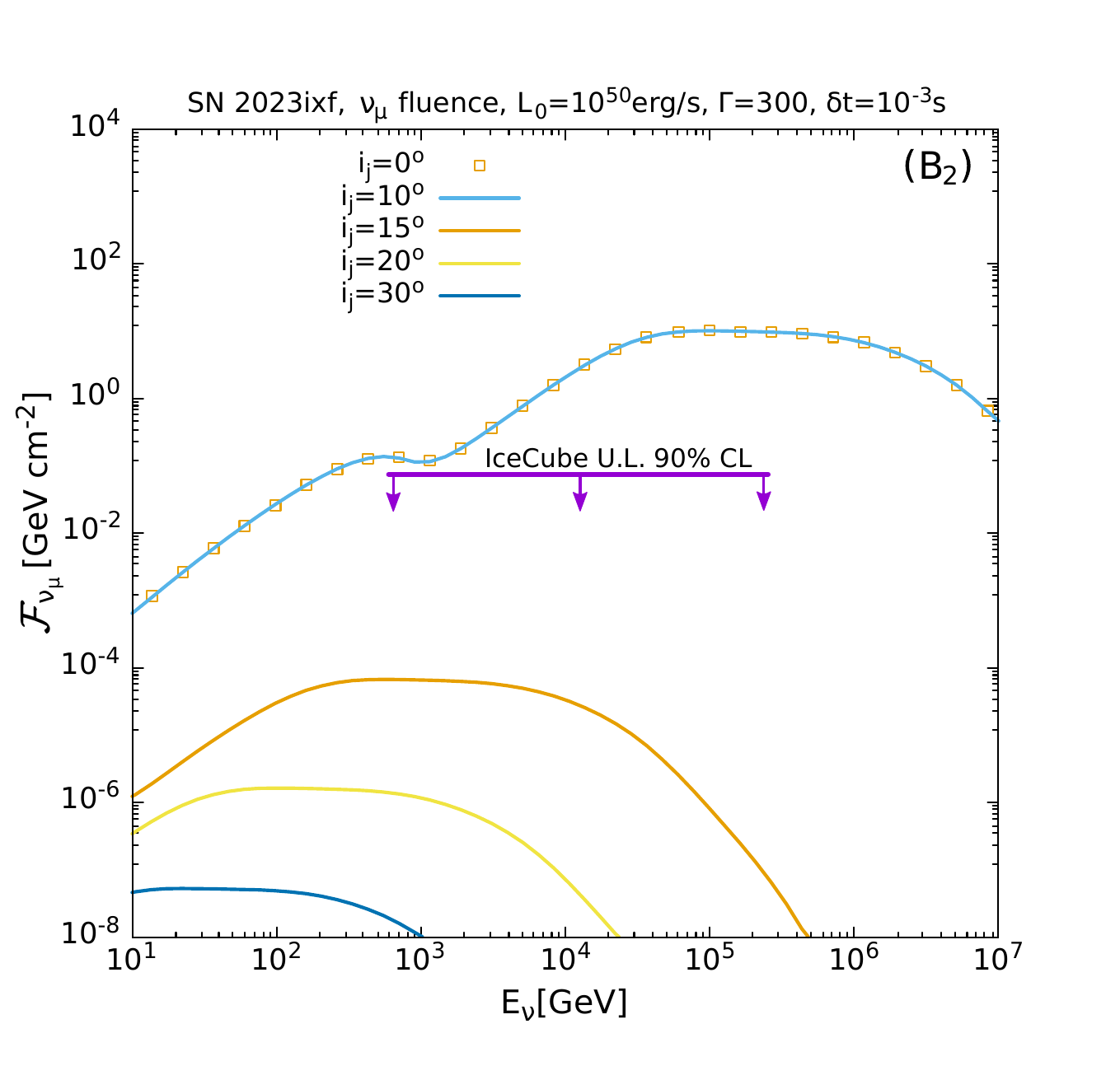} 
   \end{subfigure}     
 \caption{Different outcomes of the time-integrated flux of muon neutrinos for different values of the viewing angle $i_{\rm j}$, using the sets of parameters $B_1$ and $B_2$ in the top and bottom panels, respectively. In the top-left panel, we take $\Gamma=50$, and in the top right panel, we take $\Gamma=300$. In the bottom panels, we assume $\Gamma=75$ and $\Gamma=300$ in the left and right panels, respectively.}\label{fig6:nufluence_B1B2}
\end{figure*}

In Fig. \ref{fig5:nufluence_A1A2}, we show the obtained fluences of muon neutrinos corresponding to different viewing angles, for the sets of parameters $A_1$ and $A_2$ in the top and bottom panels, respectively. In the left panels, we assume $\Gamma=50$, while in the right panels we adopt $\Gamma=125$ for the set $A_1$ and $\Gamma=115$ for the set $A_2$. Fig. \ref{fig6:nufluence_B1B2} includes analogous plots of the fluences for the parameter sets $B_1$ and $B_2$, and in this case, $\Gamma=50$ and $\Gamma=75$ are assumed in the top-left and bottom-left panels, respectively; while $\Gamma=300$ is adopted in the right panels.

In both Fig. \ref{fig5:nufluence_A1A2} and Fig. \ref{fig6:nufluence_B1B2}, it can be seen that the outcomes corresponding to viewing angles $i_{\rm j}\lesssim \theta_{\rm op}\simeq 11^\circ$ largely overcome the level marked by the IceCube upper limit. For larger viewing angles, the corresponding fluences decrease more severely in the right panels, since for higher values of $\Gamma$ a more pronounced boosting effect is generated. 

As for the dependence on the neutrino energy obtained for $\mathcal{F}_{\nu_\mu}$, it can be seen that a somewhat flatter behaviour corresponds to the lowest values of $\Gamma$ (i.e. left panels of Figs. \ref{fig5:nufluence_A1A2},\ref{fig6:nufluence_B1B2}). This is because in those cases, $pp$ interactions play a role for low energy protons, as can be seen in Fig. \ref{fig1:p-rates}. It can also be seen that in the left panels of Fig. \ref{fig5:nufluence_A1A2}, neutrinos reach higher energies than in the right panels. This can be understood considering that in the former cases, the corresponding maximum proton energies are higher (see top panels of Fig. \ref{fig1:p-rates}) and, in addition,  pion cooling due to $\pi\gamma$ interactions is more significant for higher values of $\Gamma$ (see top panels of Fig. \ref{fig2:pi-rates}). This is partly compensated by the fact that muon synchrotron cooling is dominant for low $\Gamma$ values (see top panels of Fig. \ref{fig3:mu-rates}), and by a larger boosting effect corresponding to the high $\Gamma$ values, but still in the latter cases neutrinos reach lower energies in the observer frame.

\begin{figure*}[!h]                            
\centering
\  \centering
    \begin{subfigure}[t]{0.49\textwidth}
        \centering            \includegraphics[width=0.5\linewidth,trim= 180 30 185 35]{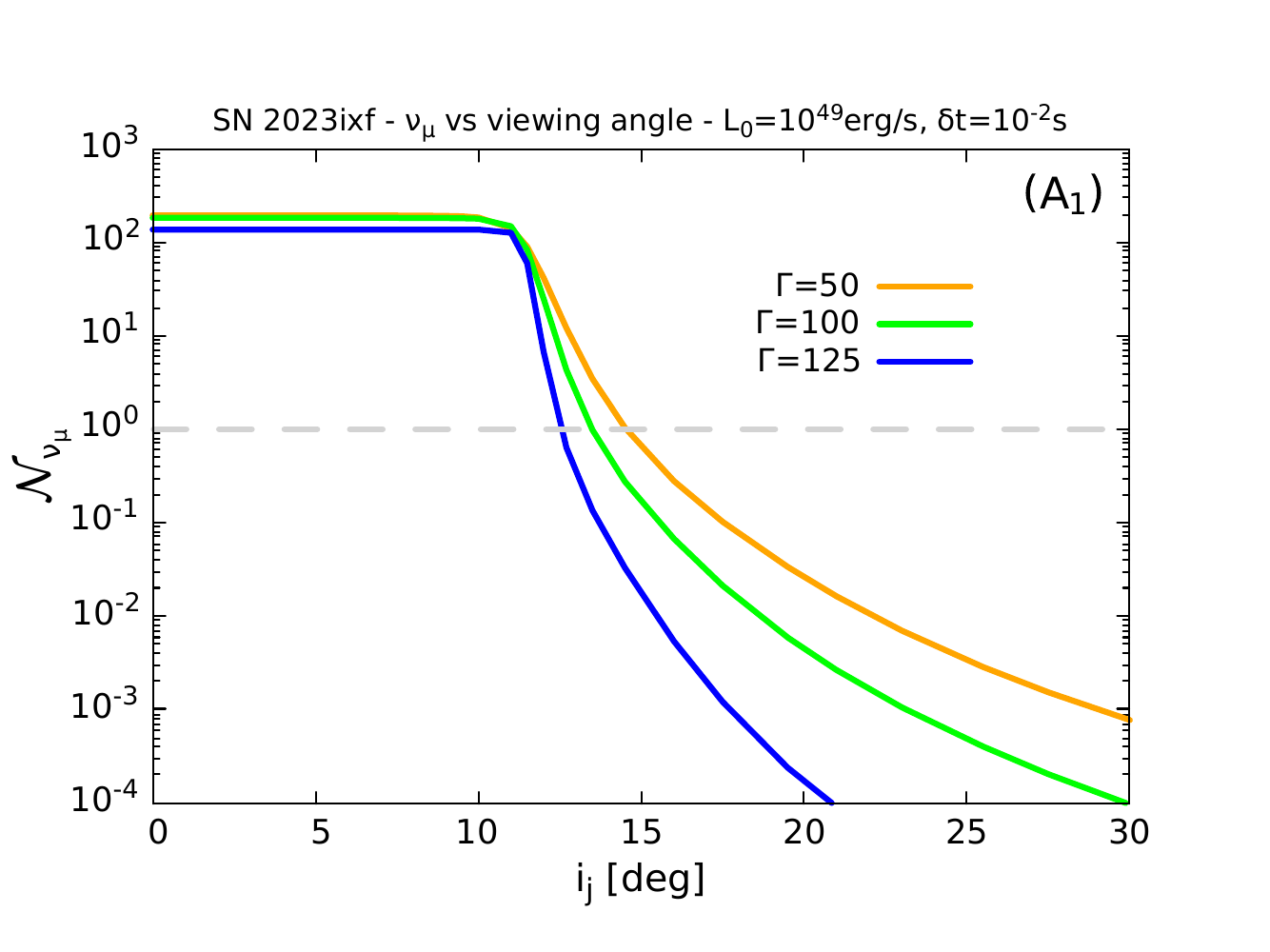} 
    \end{subfigure}
    \hfill
    \begin{subfigure}[t]{0.49\textwidth}
        \centering
    \includegraphics[width=0.5\linewidth,trim= 180 30 185 35]{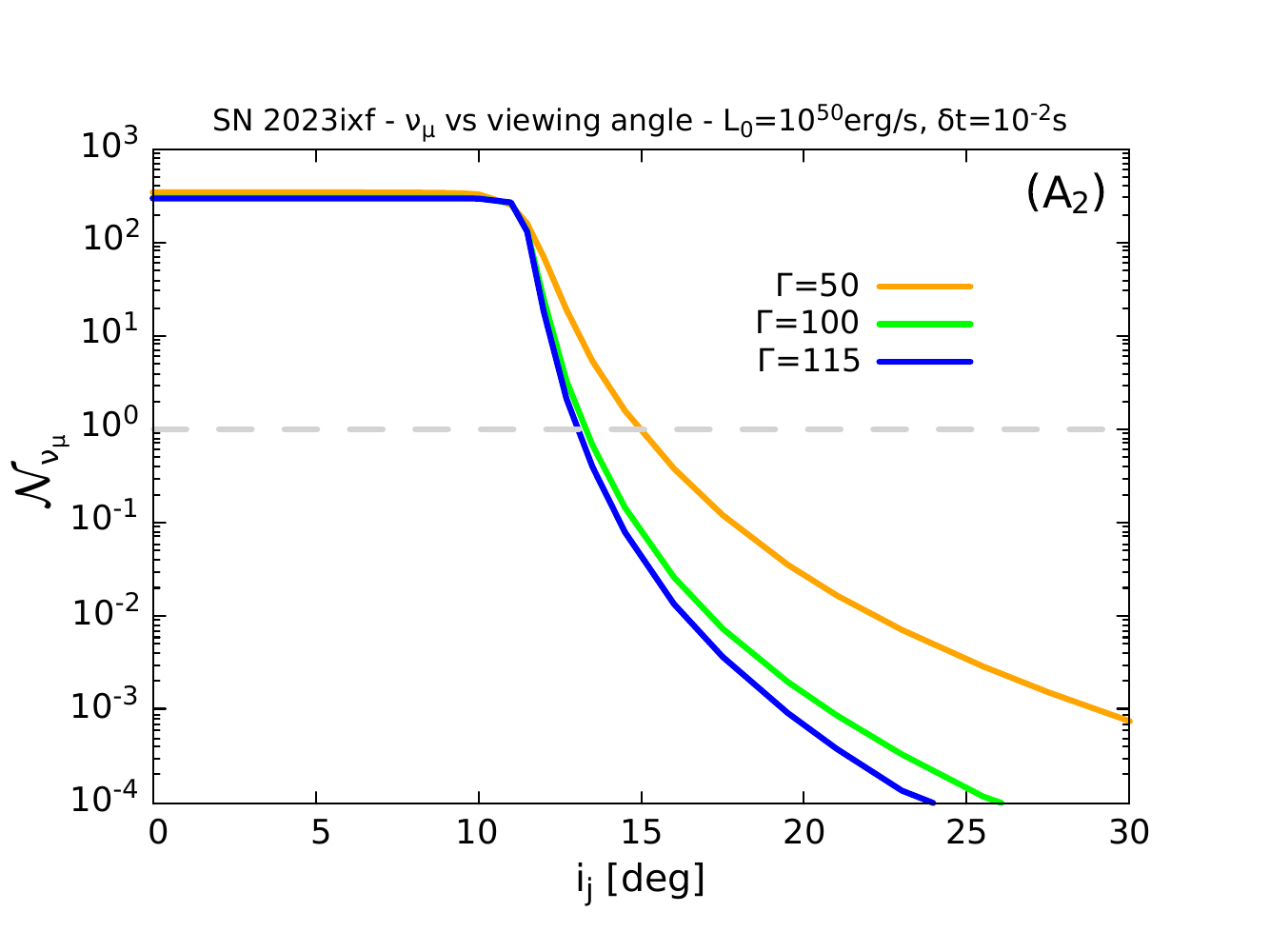} 
    \end{subfigure}
    \begin{subfigure}[t]{0.49\textwidth}
        \centering            \includegraphics[width=0.5\linewidth,trim= 180 30 185 35]{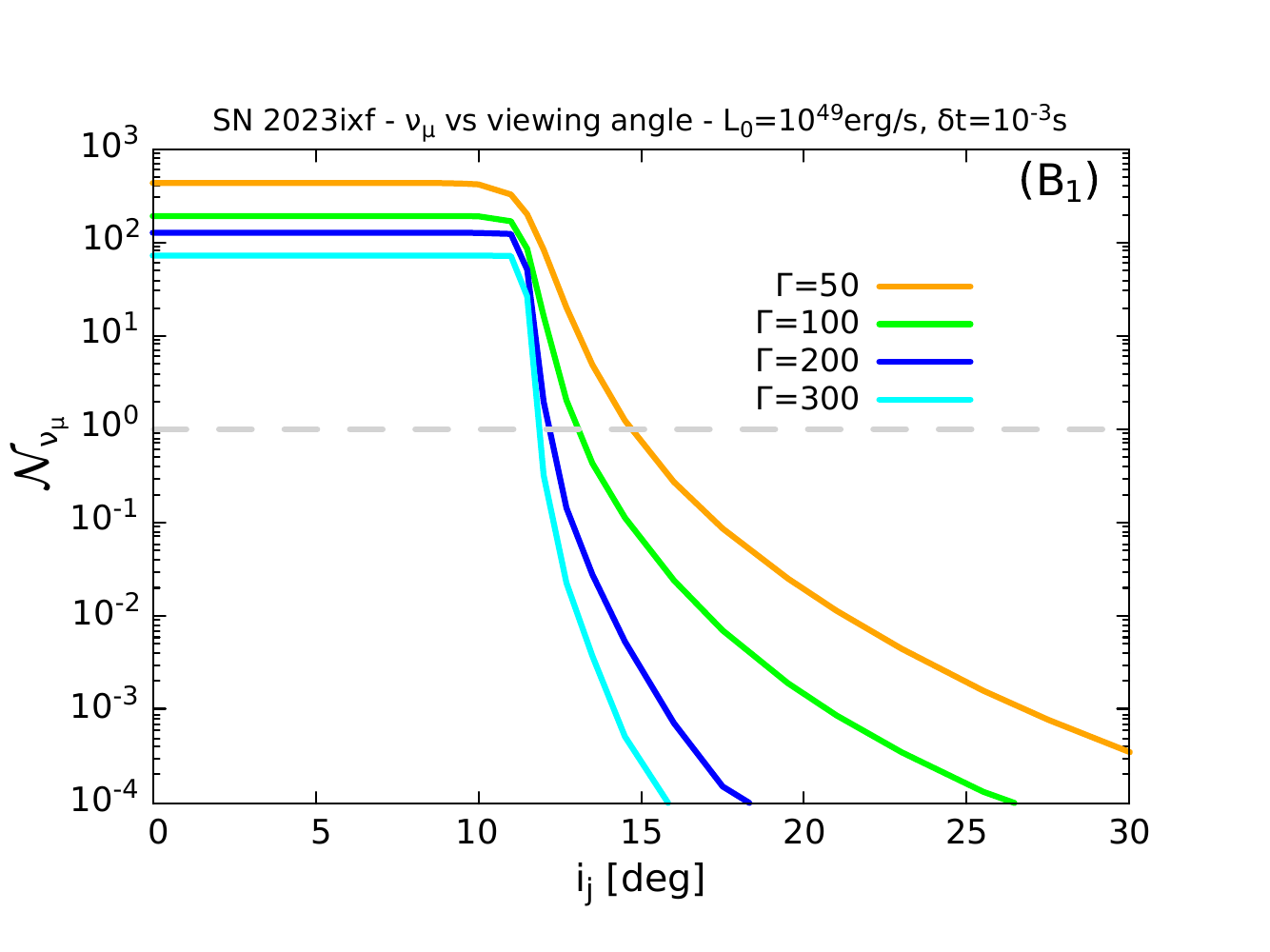} 
    \end{subfigure}
    \hfill
    \begin{subfigure}[t]{0.49\textwidth}
        \centering
        \includegraphics[width=0.5\linewidth,trim= 180 30 185 30]{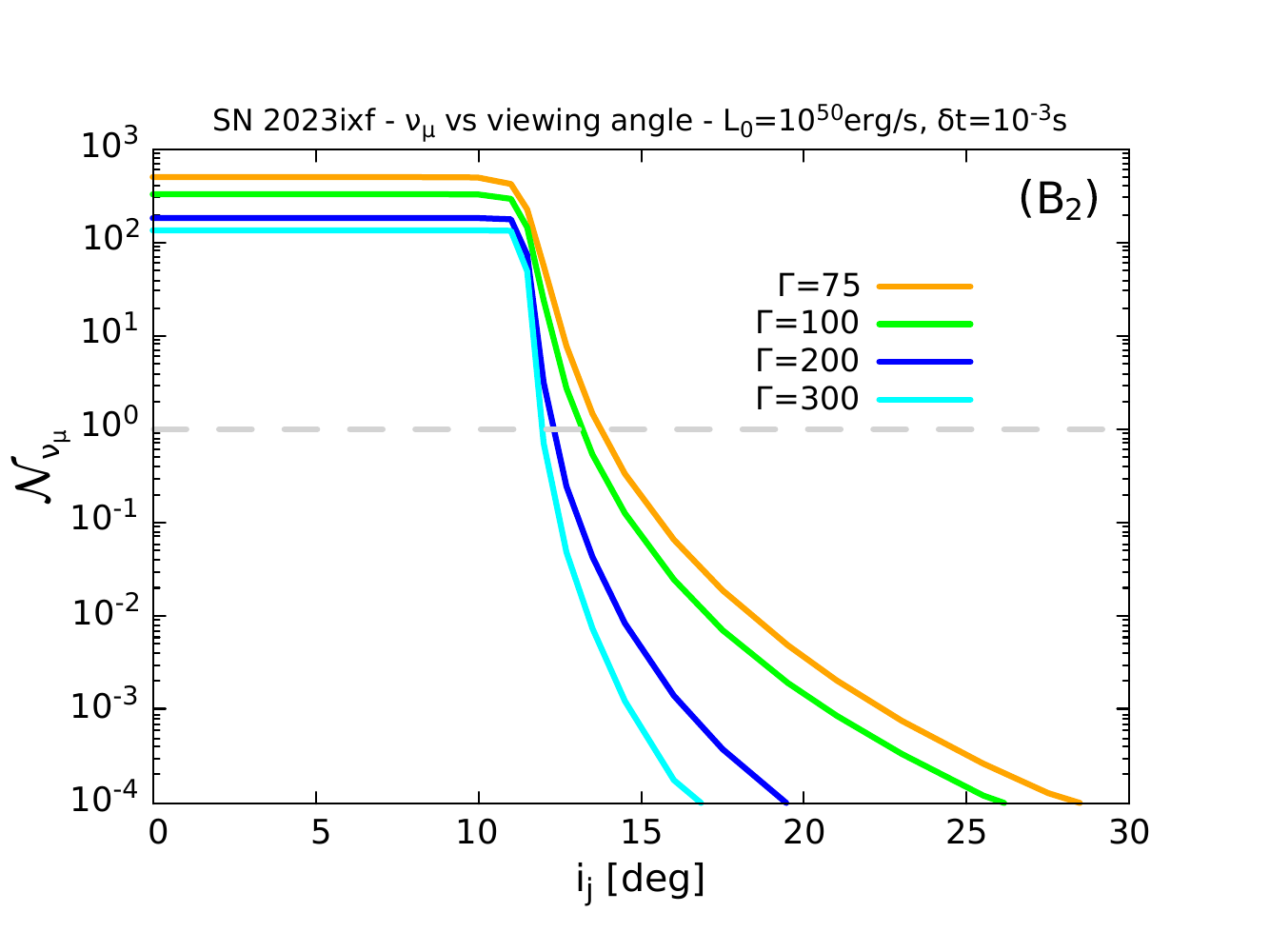} 
   \end{subfigure}     
 \caption{Number of muon neutrino events within the energy interval $600\,{\rm GeV}<E_\nu<250\,{\rm TeV}$ as a function of the viewing angle $i_{\rm j}$, and for different values of $\Gamma$. Results for the sets of parameters $A_1$ and $A_2$ appear in the top panels, while those for $B_1$ and $B_2$ are shown in the lower ones.}\label{fig7:numuevents}
\end{figure*}

We can also integrate the number of $\nu_\mu$ events for the observed jet duration $t_{\rm j}(1+z)$ and within the energy interval $600\,{\rm GeV}<E_\nu<250\,{\rm TeV}$ as
\be 
\mathcal{N}_{\nu_\mu}(i_{\rm j})=t_{\rm j}(1+z)\int_{600\,\rm GeV}^{250\,\rm TeV}dE_\nu A_{\rm eff}^{\nu_\mu}(E_\nu,\delta)\,\varphi_{\nu_\mu}(E_\nu,i_{\rm j}).
\ee
Here, we considered the neutrino effective area $A_{\rm eff}^{\nu_\mu}$ following Ref. \cite{aartsen2017} for the range of declinations $30^\circ<\delta<90^\circ$, since $\delta_{\rm M101}\simeq 54^\circ$ corresponds to the case of SN 2023ixf. The obtained results are shown in Fig. \ref{fig7:numuevents} as a function of the viewing angle $i_{\rm j}$ and for the different parameter sets considered. Again, it can be seen that the detectability sharply drops as the viewing angle exceeds $\theta_{\rm op}$. We note, in particular that for sets $B_1$ and $B_2$ the expected number of events is lower as the values of $\Gamma$ are increased significantly. This is due to a corresponding decrease in the expected fluence for low energy neutrinos, as can be seen in Fig. \ref{fig6:nufluence_B1B2}. 

\section{Discussion}\label{sec:discussion}

We have analyzed the case of neutrino production by $p\gamma$ interactions in a {hypothetical} choked jet inside a type II supernova applying a simple model to describe all the relevant cooling processes of high energy protons, pions, and muons. The basic scenario is the usually adopted one, where the target photons for the relativistic protons at the internal shocks are due to the thermalized emission generated by electrons in the jet head. We have restricted the combinations of parameters in order to satisfy the necessary radiation constraints \citep[e.g.][]{murase2013} to allow for efficient particle acceleration.

In particular, we applied the model to the specific case of the recent SN 2023ixf, obtaining the corresponding  neutrino fluences and number of events for different viewing angles. We note that this result is consistent with what is expected for an off-axis view of the jet emission in GRBs \citep{salafia2016,ahlers2019}. Regarding to this point, our approach is more general than the applied in Ref. \cite{guetta2023}, where only the case of a jet directed to the observer was considered. We still note that we obtain similar results for the number of events, considering that we assume an energy budget $L_0 t_{\rm j}=10^{53}{\rm erg}$ and a power carried by relativistic protons $L_p= \epsilon_{\rm rel}\,L_0$, with $\epsilon_{\rm rel}=0.1$.

In the context of the present model, where the jet is assumed to be conical and its bulk motion radially directed, then a closer look at the dependence of the events' number vs the viewing angle (Fig. \ref{fig7:numuevents}) reveals that even for viewing angles slightly greater than the half-opening angle, IceCube could have detected some events from SN 2023ixf. As it can be seen from Fig. \ref{fig7:numuevents}, this effect is more significant for low values of $\Gamma$, such that even for viewing angles $i_{\rm j}\lesssim 15^\circ\simeq 1.3\,\theta_{\rm op}$, the expected number of neutrino events would be $\mathcal{N}_{\nu_\mu}\gtrsim 1$.

Overall, the main conclusion after considering several plausible parameter sets for SN 2023ixf, is that the choked jet scenario can be consistent with the non-detection by IceCube if the viewing angle exceeds the jet half-opening angle ($i_{\rm j}\gtrsim\theta_{\rm op}$). Conversely, IceCube would have detected a significant number of events from SN 2023ixf ($\mathcal{N}_{\nu_\mu} \gtrsim 10^2$) if the $i_{\rm j}\lesssim\theta_{\rm op}$, and this was not the case. To summarize, either no choked jet was launched at all, or if it did, it could even have accelerated protons and generated neutrinos efficiently, but been beamed in a direction other than that of our line of sight, eluding our detection. Alternatively, the jet power could have been significantly weaker than assumed here, e.g. with a energy budget of $\lesssim 10^{50}{\rm erg}$ and even pointing in our direction, the expected neutrino signal in that case would have been undetectable by IceCube. {A similar analysis could be realized for another type II supernova that was observed very recently, SN 2024ggi  \citep{srivastav2024}. This explosion took place in the galaxy NGC 3621 at about the same distance as SN 2023ixf \citep{xiang2024}, but no associated neutrino observations have been reported.}

Hence, more observations of future nearby CCSNe would be useful to further constrain the choked jet scenario and their possible role in the explosions. In particular, with next generation, larger detectors such as IceCube-gen2 \citep{icecubegen22014}, the observability of high energy neutrino from these sources is expected to increase significantly.





\section*{Acknowledgements}
 We thank ANPCyT (Argentina) and Universidad Nacional de Mar del Plata (Argentina) for their financial support through grants PICT 2021-GRF-T1-00725 and EXA1214/24, respectively.


\bibliographystyle{elsarticle-num} 
\bibliography{paperSNnu_v2}






\end{document}